\documentclass[11pt]{article}
\usepackage{amssymb}
\usepackage{subfigure}
\usepackage{epsfig}
\usepackage{graphicx}
\usepackage{amsmath}
\usepackage{pgf,tikz}
\graphicspath{{./}{Figures/}}
\begin{document}
\title
{Homogenized boundary conditions and resonance effects\\ in Faraday cages}
\author{D.~P.~Hewett and I.~J.~Hewitt \\[2mm]
\footnotesize{Mathematical Institute, University of Oxford, UK}\\
\footnotesize{hewett@maths.ox.ac.uk, hewitt@maths.ox.ac.uk}}
\date{\today}
\maketitle
\begin{abstract}
We present a mathematical study of two-dimensional electrostatic and electromagnetic shielding by a cage of conducting wires (the so-called `Faraday cage effect'). Taking the limit as the number of wires in the cage tends to infinity we use the asymptotic method of multiple scales to derive continuum models for the shielding, involving homogenized boundary conditions on an effective cage boundary. We show how the resulting models depend on key cage parameters such as the size and shape of the wires, and, in the electromagnetic case, on the frequency and polarisation of the incident field. In the electromagnetic case there are resonance effects, whereby at frequencies close to the natural frequencies of the equivalent solid shell, the presence of the cage actually amplifies the incident field, rather than shielding it. By appropriately modifying the continuum model we calculate the modified resonant frequencies, and their associated peak amplitudes. We discuss applications to radiation containment in microwave ovens and acoustic scattering by perforated shells.
\end{abstract}
\newcommand{\done}[2]{\dfrac{d {#1}}{d {#2}}}
\newcommand{\donet}[2]{\frac{d {#1}}{d {#2}}}
\newcommand{\pdone}[2]{\dfrac{\partial {#1}}{\partial {#2}}}
\newcommand{\pdonet}[2]{\frac{\partial {#1}}{\partial {#2}}}
\newcommand{\pdonetext}[2]{\partial {#1}/\partial {#2}}
\newcommand{\pdtwo}[2]{\dfrac{\partial^2 {#1}}{\partial {#2}^2}}
\newcommand{\pdtwot}[2]{\frac{\partial^2 {#1}}{\partial {#2}^2}}
\newcommand{\pdtwomix}[3]{\dfrac{\partial^2 {#1}}{\partial {#2}\partial {#3}}}
\newcommand{\pdtwomixt}[3]{\frac{\partial^2 {#1}}{\partial {#2}\partial {#3}}}
\newcommand{\bs}[1]{\mathbf{#1}}
\newcommand{\bx}{\mathbf{x}}
\newcommand{\by}{\mathbf{y}}
\newcommand{\bd}{\mathbf{d}} 
\newcommand{\bn}{\mathbf{n}} 
\newcommand{\bP}{\mathbf{P}} 
\newcommand{\bp}{\mathbf{p}} 
\newcommand{\ol}[1]{\overline{#1}}
\newcommand{\rf}[1]{(\ref{#1})}
\newcommand{\xt}{\mathbf{x},t}
\newcommand{\hs}[1]{\hspace{#1mm}}
\newcommand{\vs}[1]{\vspace{#1mm}}
\newcommand{\eps}{\varepsilon}
\newcommand{\ord}[1]{\mathcal{O}\left(#1\right)} 
\newcommand{\oord}[1]{o\left(#1\right)}
\newcommand{\Ord}[1]{\Theta\left(#1\right)}
\newcommand{\PhiF}{\Phi_{\rm freq}}
\newcommand{\real}[1]{{\rm Re}\left[#1\right]} 
\newcommand{\im}[1]{{\rm Im}\left[#1\right]}
\newcommand{\hsnorm}[1]{||#1||_{H^{s}(\bs{R})}}
\newcommand{\hnorm}[1]{||#1||_{\tilde{H}^{-1/2}((0,1))}}
\newcommand{\norm}[2]{\left\|#1\right\|_{#2}}
\newcommand{\normt}[2]{\|#1\|_{#2}}
\newcommand{\on}[1]{\Vert{#1} \Vert_{1}}
\newcommand{\tn}[1]{\Vert{#1} \Vert_{2}}
\newcommand{\ts}{\tilde{s}}
\newcommand{\tGamma}{{\tilde{\Gamma}}}
\newcommand{\darg}[1]{\left|{\rm arg}\left[ #1 \right]\right|}
\newcommand{\bnabla}{\boldsymbol{\nabla}}
\newcommand{\dive}{\boldsymbol{\nabla}\cdot}
\newcommand{\curl}{\boldsymbol{\nabla}\times}
\newcommand{\Phixy}{\Phi(\bx,\by)}
\newcommand{\PhiOxy}{\Phi_0(\bx,\by)}
\newcommand{\dxPhixy}{\pdone{\Phi}{n(\bx)}(\bx,\by)}
\newcommand{\dyPhixy}{\pdone{\Phi}{n(\by)}(\bx,\by)}
\newcommand{\dxPhiOxy}{\pdone{\Phi_0}{n(\bx)}(\bx,\by)}
\newcommand{\dyPhiOxy}{\pdone{\Phi_0}{n(\by)}(\bx,\by)}

\newcommand{\rd}{\mathrm{d}}
\newcommand{\R}{\mathbb{R}}
\newcommand{\N}{\mathbb{N}}
\newcommand{\Z}{\mathbb{Z}}
\newcommand{\C}{\mathbb{C}}
\newcommand{\K}{{\mathbb{K}}}
\newcommand{\ri}{{\mathrm{i}}}
\newcommand{\re}{{\mathrm{e}}} 

\newcommand{\cA}{\mathcal{A}}
\newcommand{\cC}{\mathcal{C}}
\newcommand{\cS}{\mathcal{S}}
\newcommand{\cD}{\mathcal{D}}
\newcommand{\cone}{{c_{j}^\pm}}
\newcommand{\ctwo}{{c_{2,j}^\pm}}
\newcommand{\cthree}{{c_{3,j}^\pm}}

\newtheorem{thm}{Theorem}[section]
\newtheorem{lem}[thm]{Lemma}
\newtheorem{defn}[thm]{Definition}
\newtheorem{prop}[thm]{Proposition}
\newtheorem{cor}[thm]{Corollary}
\newtheorem{rem}[thm]{Remark}
\newtheorem{conj}[thm]{Conjecture}
\newtheorem{ass}[thm]{Assumption}
\newcommand{\dhc}[1]{{\color{red}{#1}}}
\newcommand{\n}{M}
\newcommand{\deltamax}{\delta_{\rm max}}
\newcommand\half{\mbox{$\frac{1}{2}$}}
\section{Introduction \label{sec:Intro}}
The Faraday cage effect is the phenomenon whereby electric fields and electromagnetic waves can be blocked by a wire mesh. The effect was demonstrated experimentally by Faraday in 1836 \cite{Faraday}, was familiar to Maxwell \cite{Maxwell}, and its practical application in isolating electrical systems and circuits is well known to modern-day engineers and physicists alike. However, somewhat surprisingly there does not seem to be a widely-known mathematical analysis quantifying the effectiveness of the shielding as a function of the basic cage properties (e.g.\ the geometry of the cage, and the thickness, shape and spacing of the wires in the mesh from which it is constructed). The recent publication \cite{ChHeTr:15} provided such an analysis for the two-dimensional electrostatic problem where the cage is a ring of $\n$ equally spaced circular wires of small radius $r\ll 1/M$ held at a common constant potential, which can be formulated as a Dirichlet problem for the Laplace equation. It was found in \cite{ChHeTr:15} that the shielding effect of such a Faraday cage is surprisingly weak: as the number of wires $\n$ tends to infinity the magnitude of the field inside the cage in general decays at best only inverse linearly in $\n$, rather than exponentially, as one might infer from certain treatments of the Faraday cage effect in the physics literature (see e.g\ \cite[Sec.~7-5]{Feynman}).

One of the key tools used in \cite{ChHeTr:15} to study the Faraday cage effect in the regime of large $\n$ was a continuum model in which the shielding effect of the discrete wires is replaced by a homogenized boundary condition on an infinitesimally thin interface between the ``inside'' and ``outside'' of the cage. Such boundary conditions can be derived by matching asymptotic expansions of the field away from the mesh with expansions in a boundary layer close to the mesh, where a multiple scales approximation can be applied (cf.\ \cite[\S5 and Appendix C]{ChHeTr:15}, and the closely related work in \cite{Delourme,DeHaJo:12,DeHaJo:13,BrChRa:15}).%

The current paper extends the analysis of \cite{ChHeTr:15} in a number of significant ways. Firstly, we explain how the homogenized boundary condition of \cite{ChHeTr:15} generalises to arbitrary wire shapes (not necessarily circular). Secondly, we investigate the `thick wire' regime in which $r = \ord{1/\n}$ (the model proposed in \cite{ChHeTr:15} is valid only for $r\ll 1/\n$ and is in general ill-posed for $r = \ord{1/\n}$.) Thirdly, we consider the analogous Neumann problem, where the interesting regime is not that of small wires, but rather small gaps between wires. 
Finally, and perhaps most significantly, we undertake a detailed study of the two-dimensional electromagnetic problem in which an external time-harmonic wave field (a solution of the Helmholtz equation) is incident on the cage. We show that, under appropriate assumptions on the wavelength and the wire radii, the leading order wave field satisfies the same homogenized boundary conditions as in the Laplace case. However, in the wave problem there is the possibility of \emph{resonance}, where the presence of the cage actually \emph{amplifies} the incident field, rather than shielding it. For the Dirichlet problem, such resonance effects are strongest in the `thick wire' regime in which $r = \ord{1/\n}$, and when the wavelength is close to (but not in general equal to) a resonant wavelength of the idealised cage in which the wire mesh is replaced by a solid shell. We show how to modify the continuum model to deal with such resonances, and use our modified model to calculate precisely the wavelength at which the maximum amplification is observed, and the associated peak amplitude, validating our predictions against numerical simulations. 

We end this introduction with some comments on related literature. Firstly, we acknowledge that there is already a substantial literature concerning the rigorous analysis of homogenization procedures for potential and scattering problems involving thin, rapidly-varying interfaces. While we do not attempt a comprehensive review, we note in particular the works \cite{CiMu:97,ClDe:13,DeHaJo:12,DeHaJo:13,Delourme,RaTa:75a,RaTa:75b,Na:08a,Na:08b,DeScSe:15}, which consider problems closely related (but different) to those studied here. Many of these studies adopt a similar multiple-scales-based approach to ours, albeit from a slightly more rigorous point of view, and some (e.g.\ \cite{ClDe:13}) derive higher order asymptotic approximations than those considered here. What sets our work apart from this literature is that we are concerned less with formulating high order approximations and proving rigorous error estimates and more with understanding the qualitative and quantitative behaviour of the leading order homogenized approximations - in particular their shielding performance - something which to date does not appear to have been studied systematically. 
Secondly, we mention \cite{Ma:14}, which treats the Dirichlet problem for a circular cage of small equally-spaced wires using the so-called ``Foldy method'' from multiple scattering theory, in which the wires act as point sources and the geometrical assumptions permit a semi-analytical solution for the associated amplitudes in terms of the discrete Fourier transform. This method appears to be closely related to the lowest order version of the Mikhlin-type numerical method used in \cite{ChHeTr:15}, higher order versions of which shall be our main source of numerical approximations for the circular wire case. 
The analysis of \cite{Ma:14} does not cover the regime $r=\ord{1/M}$ and does not treat resonance effects. 

\section{Problem formulation \label{sec:ProblemFormulation}}
\begin{figure}[t]
\begin{center}
\subfigure[\label{subfig:Gamma}]{
\includegraphics[scale=1]{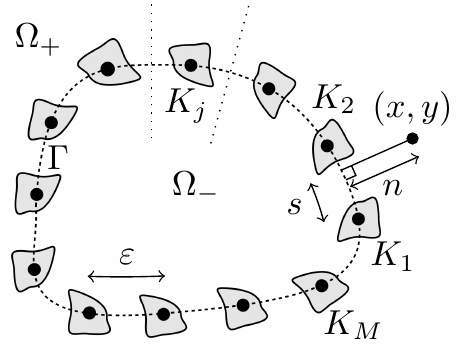}
}
\subfigure[\label{subfig:Cell}]{
\includegraphics[scale=1]{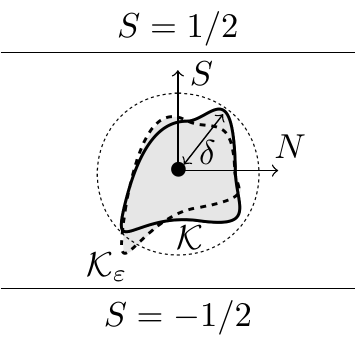}
}
\subfigure[\label{subfig:K}]{
\includegraphics[scale=1]{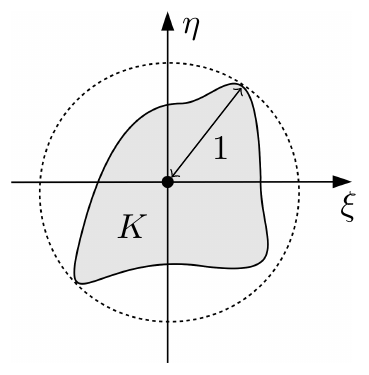}
}
\end{center}
\caption{\label{fig:geometryMOD} \subref{subfig:Gamma} Faraday cage geometry and the outer coordinates $(x,y)$ and $(n,s)$, with the curve $\Gamma$ on which the wires are centred shown as a dashed line. The dotted lines either side of the wire $K_j$ are curves of constant $s=s_j\pm \eps/2$, corresponding to the lines $S=\pm 1/2$ in the boundary layer coordinates.
\subref{subfig:Cell} The cell problem geometry and the boundary layer coordinates $(N,S)=(n/\eps,s/\eps)$, showing the scaled wire shape $\mathcal{K}$ (solid boundary; Model 2) and the perturbation $\mathcal{K}_\eps$ (dashed boundary; Model 1). 
\subref{subfig:K} The reference wire shape $K$ and the inner inner coordinates $(\xi,\eta)$. %
}
\end{figure}

Let $\Omega_-$ be a bounded simply connected open subset of the plane with smooth boundary $\Gamma=\partial\Omega_-$ and let $\Omega_+:=\R^2\setminus \overline{\Omega}_-$ denote the complementary exterior domain. For convenience we will routinely identify the $(x,y)$-plane with the complex $z$-plane, $z=x+\ri y$. 
We consider a `cage' of
$\n$ non-intersecting 
wires $\{K_j\}_{j=1}^\n$ (compact subsets of the plane, defined in more detail shortly) centred at points $\{z_j\}_{j=1}^\n$ along $\Gamma$ with constant separation (measured with respect to arc length along $\Gamma$)
\[\eps=|\Gamma|/\n,\]
where
$|\Gamma|$ is the total length of $\Gamma$; for an illustration see Figure \ref{fig:geometryMOD}\subref{subfig:Gamma}. 
We set $D:=\R^2\setminus\left(\bigcup_{j=1}^\n K_j\right)$. 

The electrostatic problem is formulated as follows. Given a compactly supported source function $f$, 
we seek a real-valued potential $\phi(z)$ satisfying
\begin{align}
\label{eqn:Laplace}
\nabla^2\phi  = f & \qquad \textrm{in }D,\\
\label{eqn:BC}
\phi=0 & \qquad \textrm{on }\partial K_j,\; j=1,\ldots,\n,\\
\label{eqn:Infinity}
\phi(z)\sim \left(\frac{1}{2\pi}\int_D f\right) \log(|z|) + \ord{1} & \qquad \textrm{as } z\to \infty.
\end{align}
Condition \rf{eqn:BC} models the fact that the wires are electrically connected, e.g.\ at infinity in the third dimension. Condition \rf{eqn:Infinity}
ensures that the cage possesses zero net charge.
We note that the formulation \rf{eqn:Laplace}-\rf{eqn:Infinity} is different (but equivalent) to that in \cite{ChHeTr:15}, where the constant term at infinity in \rf{eqn:Infinity} was zero, with $\phi$ taking an unknown (and in general non-zero) constant value on the wires. 
For completeness we also consider the Neumann problem in which \rf{eqn:BC} is replaced by 
\begin{align}
\label{eqn:BCNeu}
\pdone{\phi}{\nu}=0 & \qquad \textrm{on }\partial K_j,\; j=1,\ldots,\n,
\end{align}
where $\nu$ denotes a unit normal vector on $\partial K_j$, and $\ord{1}$ is replaced by $o(1)$ in \rf{eqn:Infinity}. While not having any obvious electrostatic application, this could represent a model for inviscid incompressible fluid flow due to a source in the presence of a cage of impermeable wires.

The time-harmonic electromagnetic problem can be formulated in terms of two complex-valued scalar fields, representing the out-of-plane components of the electric and magnetic fields respectively, both of which satisfy the Helmholtz equation
\begin{align}
\label{eqn:Helmholtz}
(\nabla^2+k^2)\phi = f \qquad \textrm{in }D,
\end{align}
for appropriate source functions $f$, where $k>0$ is the (nondimensional) wavenumber. 
The out-of-plane component of the electric field (TE mode) satisfies the Dirichlet boundary condition \rf{eqn:BC} and the out-of-plane component of the magnetic field (TM mode) satisfies the Neumann boundary condition \rf{eqn:BCNeu}. At infinity both fields are assumed to satisfy an outgoing radiation condition. These two problems also model the analogous acoustic scattering problems with sound-soft and sound-hard boundary conditions respectively.

The goal of this paper is to determine the leading order asymptotic solution behaviour of the above problems as the number of wires $\n$ tends to infinity, equivalently, as the wire separation $\eps$ tends to zero\footnote{We assume that lengths have been nondimensionalised relative to a suitable macro-lengthscale (e.g.\ the radius of the smallest circle containing $\Gamma$) so that $\eps$ is a nondimensional parameter.}.
To make this goal well-defined we need to specify how the wire size, shape and orientation should vary as $\eps\to 0$. 
In particular, in order that the wires remain disjoint as $\eps\to 0$ (so that the wires form a `cage' and not a solid shell), the wire radii must in general decrease in proportion to $\eps$ (or faster). 

We consider two different models, defining a reference wire shape either in local Cartesian coordinates aligned with $\Gamma$, or in local curvilinear coordinates that conform to $\Gamma$.  Since $\Gamma$ is smooth there is no difference between these models at leading order, but the distinction affects higher order corrections (due to the curvature of $\Gamma$) that will enter some of our calculations.  To make the definitions specific, we must introduce some further notation.

Close to $\Gamma$ we can change from Cartesian coordinates $(x,y)$ to orthogonal curvilinear coordinates $(n,s)$, such that $n$ is the distance from $(x,y)$ to the closest point on $\Gamma$ (positive/negative $n$ representing points inside $\Omega_+$ and $\Omega_-$ respectively), and $s$ is arc length along $\Gamma$ to this closest point measured counterclockwise from some reference point 
on $\Gamma$. Given a reference point $z_j$ on $\Gamma$ 
with curvilinear coordinates $(0,s_j)$, we define local curvilinear coordinates $(\tilde{n},\tilde{s})$ by $\tilde{n}=n$, $\tilde{s}=s-s_j$, and local Cartesian coordinates $(\tilde{x},\tilde{y})$ such that 
the positive $\tilde{x}$ axis is aligned to the positive $\tilde{n}$ axis at $z_j$.
Explicitly, $\tilde{x}+\ri \tilde{y}=\re^{-\ri \theta_j}(z-z_j)$, where $\theta_j$ is the counter-clockwise angle from the positive $x$ axis to the outward normal vector to $\Gamma$ at $z_j$. 
To convert between these coordinate systems there exists a diffeomorphism $F_j: (-n_j,n_j)\times(-\eps/2,\eps/2) \to U_j$, where $U_j$ is an open neighbourhood of $z_j$ and $n_j>0$ is a constant, such that $(\tilde{x},\tilde{y}) = F_j(\tilde{n},\tilde{s})$ (see Appendix \ref{sec:higherorder}).

We are now ready to specify the wire geometries and their dependence on $\eps$. 
For both models, we assume a fixed reference wire shape $K$;
a compact subset of the plane for which the smallest closed disc containing $K$ has radius one and is centred at the origin (see Figure \ref{fig:geometryMOD}\subref{subfig:K}). 

In Model 1 we define a wire $K_j$ of radius $r>0$ centred at $z_j$ by the formula $K_j=rK$ in the $(\tilde{x},\tilde{y})$ coordinate system, which in the original $z$ coordinates gives
\begin{align}
\label{eqn:WireDef1} 
K_j = z_j + \re^{\ri\theta_j}(rK).
\end{align}
In Model 2 we use the same formula $K_j=rK$ but interpreted in the $(\tilde{n},\tilde{s})$ coordinate system, which in the original $z$ coordinates gives
\begin{align}
\label{eqn:WireDef2} 
K_j = z_j + \re^{\ri\theta_j} F_j(rK).
\end{align}

\begin{figure}[t]
\begin{center}
\subfigure[\label{subfig:circles}]{
\includegraphics[scale=1]{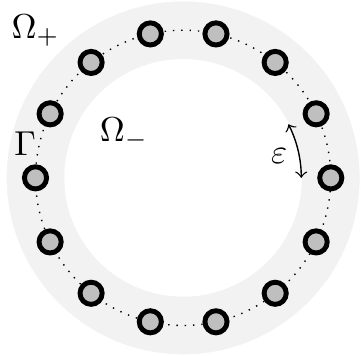}
}
\hs{3}
\subfigure[\label{subfig:linesegments}]{
\includegraphics[scale=1]{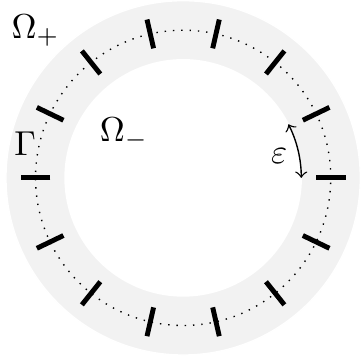}}
\hs{3}
\subfigure[\label{subfig:arcsofcircles}]{
\includegraphics[scale=1]{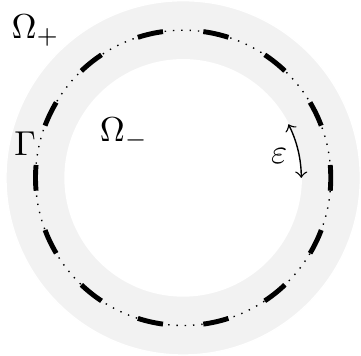}}
\end{center}
\caption{\label{fig:geometry}Faraday cage geometries for $\Gamma$ a circle.  In \subref{subfig:circles} the reference wire shape $K$ is a closed disk, in \subref{subfig:linesegments} it is the line segment $[-1,1]$, and in \subref{subfig:arcsofcircles} it is the line segment $[-i,i]$.
Model 1 is used in \subref{subfig:circles} (the wires would be slightly deformed disks under Model 2), there is no difference between the two wire models in \subref{subfig:linesegments}, and Model 2 is used in \subref{subfig:arcsofcircles} (the wires would be tangential line segments under Model 1, rather than circular arcs).}
\end{figure}

Examples are illustrated in Figure \ref{fig:geometry}.
The rationale for considering both wire models is that 
Model 1 is the more natural from a physical point of view as the wire shape is independent of $r$ in the original Cartesian coordinate system, while Model 2 is simpler from a mathematical point of view as the wire shape is independent of $r$ in the curvilinear coordinates in which we derive our homogenized boundary conditions (see \S\ref{sec:EquivBCs}). In many aspects of our analysis the two models produce the same results. But for some problems requiring higher-order boundary layer expansions they may produce different results. 

In order that the wires remain disjoint as $\eps\to 0$
we assume that the wire radius $r$ satisfies
\[ r = \delta \eps,\]
where $0<\delta=\delta(\eps)<\deltamax$ and $\deltamax=\ord{1}$ is the critical scaling that gives rise to touching wires in the limit as $\eps\to 0$. For example, $\deltamax=1/2$ for both the case of circular wires, when $K$ is the unit disk, cf.\ Figure \ref{subfig:circles}, and the case of tangential line segments, cf.\ Figure \ref{subfig:arcsofcircles}. 
An exceptional case where no such $\deltamax$ exists is that of line-segment wires perpendicular to $\Gamma$, when $K$ is the interval $[-1,1]$, cf.\ Figure \ref{subfig:linesegments}.  
Note in particular that a fixed value for $\delta$ corresponds to the wires taking up a fixed total fraction of the length of $\Gamma$, as the number of wires is increased. 

Our aim is to describe both qualitatively and quantitatively how the asymptotic solution behaviour of the boundary value problems depends on the reference wire shape $K$, the scaling parameter $\delta$, and in the electromagnetic case the wavenumber $k$. In doing so we generalise the analysis of \cite{ChHeTr:15}, which considered only the electrostatic case, with circular wires and the small wire regime $\delta \ll 1$. 

\section{Homogenized boundary conditions \label{sec:EquivBCs}}
In the limit $\eps\to 0$ we look for outer approximations in $\Omega_\pm$ of the form
\begin{align}
\label{eqn:2t0}
\phi(x,y)= \phi^\pm_0(x,y) + \eps\phi^\pm_1(x,y) + \ord{\eps^2} \qquad \textrm{in } \Omega_\pm,
\end{align}
where, assuming that both $f$ and $k$ are $\ord{1}$, the functions $\phi_0^\pm$ satisfy either \rf{eqn:Laplace} or \rf{eqn:Helmholtz} (as appropriate) in $\Omega_\pm$, with $\phi_1^\pm$ satisfying the homogeneous version of the same equation. Our aim is to derive homogenized boundary conditions for these functions on the interface $\Gamma$, by matching with an appropriate boundary layer solution in a region of width $\ord{\eps}$ around $\Gamma$ in which a multiple scales approximation can be applied.

We first note that 
in the curvilinear coordinates $(n,s)$
the Laplacian is \cite[(6.2.4)]{BaBu:91}
\begin{align}
\label{eqn:Laplacian}
\nabla^2 &= \frac{1}{1+\kappa n}\pdone{}{s}\left[\frac{1}{1+\kappa n}\pdone{}{s} \right] + \frac{\kappa}{1+\kappa n}\pdone{}{n} + \pdtwo{}{n},
\end{align}
where $\kappa=\kappa(s)$ is the local (signed) curvature of $\Gamma$ at the point $(0,s)$, defined with respect to a counterclockwise parametrisation. 
We introduce boundary layer variables $(N,S)$ via $(n,s)=(\eps N,\eps S)$. 
The inner limits of the outer solutions correct to $\ord{\eps}$ are found by rewriting \rf{eqn:2t0} with $n$ replaced by $\eps N$ and re-expanding, %
giving
\begin{align}
\label{eqn:InnerLimit}
\phi^\pm_0(0,s) + \eps\left( N\pdone{\phi^\pm_0}{n}(0,s)+ \phi^\pm_1(0,s) \right) + \ord{\eps^2}, 
\end{align}
with the $+$ and $-$ signs for the cases $N>0$ and $N<0$ respectively. 

In the boundary layer we look for a solution in multiple-scales form
\begin{align}
\label{}
\phi(n,s)=\Phi(N,S;s),
\end{align}
where $\Phi(N,S;s)$ is assumed to be $1$-periodic in the fast tangential variable $S$.  
To determine the equation satisfied by $\Phi(N,S;s)$ we replace $\pdonetext{}{n}$ by $\eps^{-1}\pdonetext{}{N}$ and $\pdonetext{}{s}$ by $\eps^{-1}\pdonetext{}{S} + \pdonetext{}{s}$ in \rf{eqn:Laplacian} and expand. The leading order result, for both the electrostatic and the wave problems (assuming $k=\ord{1}$), 
and for both wire Models 1 and 2,
is
\begin{align}
\label{eqn:innerOneTerm}
\pdtwo{\Phi}{N} + \pdtwo{\Phi}{S}  + \ord{\eps} = 0 \qquad \textrm{in }\mathcal{B},
\end{align}
where $\mathcal{B}=\{(N,S):|S|<1/2\}\setminus \mathcal{K}$, and $\mathcal{K}=\delta K$ (see Figure \ref{subfig:Cell}).  Periodicity requires
\begin{align}
\label{eqn:periodicity}
\Phi(N,-1/2;s)=\Phi(N,1/2;s),
\end{align}
and the conditions on $\partial\mathcal{K}$ are homogenous Dirichlet or Neumann conditions, as appropriate.  The solution is required to match with the outer solution in (\ref{eqn:InnerLimit}) as $N \to \pm \infty$.

A more detailed derivation of this boundary-layer problem is given in Appendix \ref{sec:higherorder}, where we also continue the expansion to $\ord{\eps}$.  The analysis of the $\ord{\eps}$ terms is more involved for Model 1 than for Model 2, because we have to account for the curvature of $\Gamma$ and its distorting effect on the wire shape in the $(N,S)$ coordinates (shown by $\mathcal{K}_{\eps}$ in Figure \ref{subfig:Cell}). This distortion can be neglected in the leading order problem above (and does not arise in Model 2); consequently, we leave these awkward details to the appendix.

\subsection{Dirichlet boundary conditions}

In the case of Dirichlet boundary conditions the leading order behaviour of the boundary layer solution $\Phi(N,S;s)$ with linear behaviour as $N\to\pm \infty$ (required for matching) can be written as  
\begin{align}
\label{eqn:originalinner}
\Phi(N,S;s) = 
\eps \Big( A^+(s)\Phi^+(N,S) + A^-(s)\Phi^-(N,S)\Big)+ \ord{\eps^2},
\end{align}
where the functions $\Phi^\pm(N,S)$ satisfy the following canonical cell problems 
(cf.\ Figure \ref{fig:geometryMOD}\subref{subfig:Cell}):
\begin{align}
\label{eqn:cell1}
\pdtwo{\Phi^{\pm}}{N} + \pdtwo{\Phi^{\pm}}{S}=0 & \qquad\textrm{in } \mathcal{B},\\
\Phi^\pm(N,-1/2)=\Phi^\pm(N,1/2), & \\
\Phi^\pm =0 &\qquad \textrm{on } \partial\mathcal K,\\
\Phi^+(N,S)\sim
\begin{cases}
N + \sigma_+, & N\to\infty,\\
\tau_+, & N\to-\infty,\\
\end{cases}
& \qquad
\Phi^-(N,S)\sim
\begin{cases}
\tau_-, & N\to\infty,\\
-N+\sigma_-, & N\to-\infty.\\
\end{cases}
\label{eqn:cell2}
\end{align}

For any given reference wire shape $K$ and scaled radius $\delta$ one must solve \rf{eqn:cell1}-\rf{eqn:cell2}, either analytically or numerically, to determine the far-field constants $\sigma_\pm$ and $\tau_\pm$; some specific examples are studied in Appendix \ref{sec:cells}. 
We note that if $K$ is symmetric in $\xi$ then
\begin{align}
\label{eqn:PhiPMSymmetry}
\Phi^-(N,S) = \Phi^+(-N,S), \qquad \sigma_+=\sigma_-, \quad \tau_+=\tau_-.
\end{align}
Furthermore, we note that if $\delta\ll 1$ the scaled wire $\mathcal K$ effectively acts as a point sink in the cell domain,
and a generalisation of the argument in \cite[\S C]{ChHeTr:15} proves that, outside an $\ord{\delta}$ neighbourhood of $\mathcal{K}$, %
\begin{align}
\label{eqn:BLSolnDeltaSmall}
\Phi^+(N,S) \sim \frac{1}{2\pi} \Re \left\{ \pi Z + \log \left(2 \sinh \pi Z\right) + \log\frac{1}{2\pi\delta}+ a_0 \right\},
\qquad Z = N + i S,
\end{align}
where the $K$-dependent constant $a_0$ satisfies $a_0=\lim_{\varrho\to\infty}(\psi-\log{\varrho})$, where $\psi$ is the unique solution of Laplace's equation in $\R^2\setminus K$ such that $\psi=0$ on $\partial K$ and $\psi\sim \log{\varrho}+\ord{1}$ as $\varrho\to\infty$, where $\varrho=\sqrt{\xi^2+\eta^2}$.  This constant is related to the logarithmic capacity of $K$, $c(K)$, by $a_0 = -\log c(K)$ \cite{Ro:95}.   For $K$ the unit disc, $a_0=0$; for $K$ a line segment of length $2$, $a_0 = \log{2}$ (for details see Appendix \ref{sec:cells}). 
From \rf{eqn:BLSolnDeltaSmall} it follows that
\begin{align}
\label{eqn:MuNuDeltaSmall}
\sigma_\pm,\tau_\pm \sim \frac{1}{2\pi}\left(\log{\frac{1}{2\pi\delta}} + a_0 \right) + \ord{\delta}, \qquad \delta\to 0.
\end{align}

Having extracted the far-field constants $\sigma_\pm,\tau_\pm$ from the solutions of (\ref{eqn:cell1})-(\ref{eqn:cell2}), matching the linear behaviour of (\ref{eqn:originalinner}) with that of (\ref{eqn:InnerLimit}) gives
\begin{align}
\label{}
A^+(s) = \pdone{\phi^+_0}{n}(0,s), 
\qquad
A^-(s) = -\pdone{\phi^-_0}{n}(0,s),
\end{align}
and matching constant terms then requires
\begin{align}
\label{eqn:MatchPlus2}
\eps\sigma_+\pdone{\phi^+_0}{n} -\eps\tau_-\pdone{\phi^-_0}{n} = \phi_0^+ + \eps\phi^+_1 & \qquad \textrm{on }\Gamma,\\
\label{eqn:MatchMinus2}
 \eps\tau_+\pdone{\phi^+_0}{n} -\eps\sigma_-\pdone{\phi^-_0}{n} = \phi_0^- + \eps \phi^-_1  &\qquad  \textrm{on }\Gamma.
\end{align}
To proceed further we must consider the magnitude of the parameters $\sigma_\pm, \tau_\pm$, which depend on the size of $\delta$ (see Figure \ref{fig:inner1} for example). There are essentially three different regimes to consider.

\textbf{Thick wires ($\delta=\ord{1}$).} 
If $\delta$ is strictly $\ord{1}$ then 
$\sigma_\pm, \tau_\pm$ are $\ord{1}$. 
Hence, at $\ord{1}$ in (\ref{eqn:MatchPlus2}) and (\ref{eqn:MatchMinus2}),
\begin{align}
\label{eqn:ContZero}
 \phi_0^+ =  \phi_0^- = 0  &\qquad \textrm{on }\Gamma,
\end{align}
so the leading order solution is that for a perfectly reflecting (Dirichlet) boundary at $\Gamma$. %
At $\ord{\eps}$, %
\begin{align}
\label{eqn:Hom1}
\phi_1^+ = \sigma_+\pdone{\phi^+_0}{n} - \tau_-\pdone{\phi^-_0}{n}  &\qquad \textrm{on }\Gamma, \\
\label{eqn:Hom2}
\phi_1^- = \tau_+\pdone{\phi^+_0}{n} - \sigma_-\pdone{\phi^-_0}{n}  &\qquad \textrm{on }\Gamma.
\end{align}

\textbf{Thin wires ($\delta\ll 1$).}
If $\delta\ll 1$ then $\sigma_\pm, \tau_\pm\gg 1$ (cf.\ \rf{eqn:MuNuDeltaSmall}). In particular, there is a distinguished scaling in which $\sigma_\pm, \tau_\pm = \ord{1/\eps}$, which requires $\delta$ to be exponentially small with respect to $1/\eps$, i.e.\ $\delta = \ord{\re^{-c/\eps}}$ for some $c>0$. (This is essentially the same scaling as that considered in \cite{RaTa:75a,RaTa:75b,CiMu:97} in a related context.)  
Suppose that we are in this regime, with $\sigma_\pm, \tau_\pm \sim \tilde{a}_1/\eps + \tilde{a}_0$ for some $\tilde{a}_1,\tilde{a_0}$. (E.g.\ if $\delta\sim A\re^{-c/\eps}$ then $\tilde{a}_1=c/(2\pi)$ and $\tilde{a_0}=(\log(1/(2\pi A))+a_0)/(2\pi)$.) Then at $\ord{1}$ in \rf{eqn:MatchPlus2}-\rf{eqn:MatchMinus2} we find that $\phi_0$ is continuous across $\Gamma$ (i.e., $\phi_0^+=\phi_0^-$) and satisfies 
\begin{align}
\label{eqn:Hom30}
\left[ \pdone{\phi_0}{n}\right] = \tilde{\alpha} \phi_0 \qquad \textrm{on }\Gamma,
\end{align}
where $\left[ \pdonetext{\phi_0}{n}\right]=\pdonetext{\phi^+_0}{n} - \pdonetext{\phi^-_0}{n}$ and $\tilde{\alpha} = 1/\tilde{a}_1$.
Higher order matching not detailed here (requiring higher order expansion of the boundary layer problem as in Appendix \ref{sec:higherorder}) reveals that the two-term approximation $\phi_0+\eps\phi_1$ is also continuous across $\Gamma$ and satisfies a similar condition,
\begin{align}
\label{eqn:Hom3}
\left[ \pdone{\phi_0}{n}+ \eps\pdone{\phi_1}{n}\right] = \alpha \left(\phi_0+\eps \phi_1\right) \qquad \textrm{on }\Gamma,
\end{align}
where 
$\alpha = 1/(\tilde{a}_1+\eps \tilde{a}_0)$. 
Recalling \rf{eqn:MuNuDeltaSmall}, we can express $\alpha$ in terms of $\delta$ as
\begin{align}
\label{eqn:alpha}
\alpha = \frac{2\pi}{\eps\left( \log{1/(2\pi\delta)} + a_0\right)},
\end{align}
which, in the special case of circular wires (for which $a_0=0$) agrees with the effective boundary condition derived in \cite[\S C]{ChHeTr:15}. Note that  \rf{eqn:Hom3} is valid for the two-term approximation $\phi_0+\eps\phi_1$; hence in this distinguished scaling the boundary condition derived in \cite[\S C]{ChHeTr:15} gives the solution correct to $\ord{\eps}$, not just to $\ord{1}$. 
This explains the excellent agreement observed in \cite{ChHeTr:15} between numerical solutions of the electrostatic problem and solutions of the outer problem subject to (\ref{eqn:Hom3}), even when $\delta$ is not particularly small. 
We also note, however, that as $\delta$ increases, there may (depending on the value of $a_0$) come a point at which $\alpha$ blows up to infinity; precisely, this occurs at the critical value $\delta_\infty = \re^{-a_0}/(2\pi)$ (for circular wires $\delta_\infty=1/(2\pi)\approx 0.16<\delta_{\rm max}=1/2$). For $\delta>\delta_\infty$, $\alpha$ is negative and the resulting outer problem may be ill-posed (see later).  But of course for such large values of $\delta$ we are outside of this `thin-wire' regime and the conditions \rf{eqn:ContZero}--\rf{eqn:Hom2} should be used instead of (\ref{eqn:Hom3}).

\textbf{Very thin wires ($\delta\ll \ord{\re^{-c/\eps}}$).}
If $\delta \ll \ord{\re^{-c/\eps}}$ for every $c>0$, then $\sigma_\pm,\tau_\pm\gg 1/\eps$ and $\alpha\ll 1$, so that the leading order outer solution $\phi_0$ is just the free field solution of \rf{eqn:Laplace} or \rf{eqn:Helmholtz}, i.e.\ that which would exist without the presence of the cage, and there is no shielding.

\subsection{Neumann boundary condition}
In the case of Neumann boundary conditions the requirement of linearity as $N\to\pm\infty$ means that the leading order boundary layer solution can be expressed as 
\begin{align}
\label{eqn:BLSolnNeu}
\Phi(N,S;s) = A_0(s) + \eps \Big(A_1(s)  +  B_1(s) \Psi(N,S)\Big)  + \ord{\eps^2},
\end{align}
where $\Psi(N,S)$ satisfies the canonical cell problem
\begin{align}
\label{eqn:Ncell1}
\pdtwo{\Psi}{N}+\pdtwo{\Psi}{S}=0 &\qquad \textrm{in }\mathcal B,\\
\Psi(N,-1/2)=\Psi(N,1/2),  &\qquad\textrm{on } S=\pm 1/2,\\
\pdone{\Psi}{\nu} =0 &\qquad\textrm{on } \partial\mathcal K,\\
\Psi(N,S) \sim
N \pm \lambda, &\qquad N\to \pm \infty, 
\label{eqn:Ncell2}
\end{align}
in which the constant $\lambda$ is determined as part of the solution.  This problem also appears elsewhere in acoustics and fluid flow; it is sometimes referred to as a `blockage problem', and the constant $\lambda$ as a `blockage coefficient' \cite{Tu:75,Cr:11,Cr:16}. Example solutions for $\Psi(N,S)$ and $\lambda$ are given in Appendix \ref{sec:cells}. 

Matching linear terms between \rf{eqn:InnerLimit} and (\ref{eqn:BLSolnNeu}) gives that 
\begin{align}
\label{eqn:neumannmatch2}
B_1(s) = \pdone{\phi^+_0}{n}=\pdone{\phi^-_0}{n}  \qquad \textrm{on }\Gamma,
\end{align}
so the gradient of the outer problem is continuous across $\Gamma$.  Matching constant terms then gives
\begin{align}
\label{eqn:neumannmatch1}
A_0(s)+\eps A_1(s) \pm \lambda \eps \pdone{\phi_0}{n} = \phi_0^\pm + \eps \phi_1^\pm   \qquad \textrm{on }\Gamma.
\end{align}

As in the Dirichlet case, to interpret \rf{eqn:neumannmatch1} we must consider the magnitude of $\lambda$, which depends on both $K$ and $\delta$.  The interesting limit in which $\lambda$ is large is now not $\delta\to 0$, but rather $\delta\to \deltamax$, where $\deltamax$ is the critical value of $\delta$ for which $\partial\mathcal K$ touches the cell walls $S=\pm 1/2$. (Recall that $\deltamax=1/2$ for $K$ a disk.) When $\deltamax-\delta\ll 1$ we have $\lambda \gg 1$.  We consider separately the cases $\lambda = \ord{1}$, $\lambda = \ord{1/\eps}$, and $\lambda \gg 1/\eps$.

\textbf{Large gaps ($\deltamax-\delta=\ord{1}$).} 
In this case $\lambda = \ord{1}$, and (\ref{eqn:neumannmatch1}) implies that
\begin{align}
\label{}
A_0(s) = \phi_0^+ = \phi_0^-  \qquad \textrm{on }\Gamma,
\end{align}
so that, recalling \rf{eqn:neumannmatch2}, both $\phi_0$ and its normal derivative are continuous across $\Gamma$.  Hence the leading order outer solution is just the free field solution of \rf{eqn:Laplace} or \rf{eqn:Helmholtz}, and there is no shielding.

\textbf{Small gaps ($\deltamax-\delta\ll 1$).} In this case $\lambda \gg 1$.  We first consider the case $\lambda = \ord{1/\eps}$ and suppose $\lambda \sim \tilde{b}_{1}/\eps + \tilde{b}_0$. For the case of circular wires, this would occur if $\half -\delta = \ord{\eps^2}$; for line segments it would require $\half-\delta = \ord{e^{-c/\eps}}$ for some $c>0$ (see Appendix \ref{sec:cells}). 
Matching the constant terms then gives
\begin{align}
\label{}
A_0(s) \pm \tilde{b}_{1} B_1(s) = \phi_0^\pm  \qquad \textrm{on }\Gamma,
\end{align}
which together with (\ref{eqn:neumannmatch2}), and defining $\tilde{\beta}=2 \tilde{b}_{1}$ and $[\phi_0]$ = $\phi_0^+-\phi_0^-$, implies
\begin{align}
\label{eqn:neumannmatch3}
\big[\phi_0 \big] = \tilde{\beta} \pdone{\phi_0}{n}  \qquad \textrm{on }\Gamma.
\end{align}

For completeness we quote the higher order matching conditions, obtained using the results in Appendix \ref{sec:higherorder}, 
\begin{align}
\label{eqn:higherorderneumann1}
\left[\pdone{\phi_1}{n} \right] = 2 \kappa(\tilde{\mu}-\check{\mu}) \pdone{\phi_0}{n} + 2 \hat{\mu} \frac{\partial^2 \phi_0}{\partial n \partial s} - 2\check{\mu} \pdtwo{\phi_0}{n} \qquad \textrm{on }\Gamma,
\end{align}
\begin{align}
\label{eqn:higherorderneumann2}
\big[\phi_1\big] = 2 b_0 \pdone{\phi_0}{n} + b_1 \left( \pdone{\phi_1^+}{n}+\pdone{\phi_1^-}{n}\right)\qquad \textrm{on }\Gamma,
\end{align}
where $\tilde{\mu}$, $\hat{\mu}$ and $\check{\mu}$ are constants determined from the higher order boundary-layer solutions.  These depend on the precise shape of the wires.

Rather than embarking on a detailed study of different cases, we concentrate on the case that is perhaps of most interest for this small-gap situation; namely, when the wires form a perforated shell around $\Gamma$ (cf.\ Figure \ref{subfig:arcsofcircles}).  This corresponds to tangential line segments (i.e.\ $K=[-\ri,\ri]$) under Model 2, for which we find $\tilde{\mu} = \check{\mu} = \hat{\mu} = 0$, and $\lambda\sim-(1/\pi)(\log{\pi(1/2-\delta)})$ (Appendix \ref{sec:cells}).   In this case (\ref{eqn:higherorderneumann1}) and (\ref{eqn:higherorderneumann2}) combine with (\ref{eqn:neumannmatch3}) to give 
\begin{align}
\label{eqn:neumannmatch4}
\big[\phi_0 + \eps \phi_1 \big] = \beta \left( \pdone{\phi_0}{n} + \eps \pdone{\phi_1}{n}  \right) \qquad \textrm{on }\Gamma,
\end{align}
where $\beta = 2(\tilde{b}_1 + \eps \tilde{b}_0)$.
If $\delta=1/2-A\re^{-c/\eps}$ then $\beta= 2c/\pi - 2 \eps \log(\pi A)/\pi $.  
There is a duality between (\ref{eqn:neumannmatch4}) and condition (\ref{eqn:Hom3}) that holds in the Dirichlet case, although we note that for more general wire shapes (\ref{eqn:neumannmatch4}) may become more complicated.

\textbf{Very small gaps ($\deltamax-\delta\ll 1$).}  In the case that $\lambda \gg \ord{1/\eps}$, matching constant terms in (\ref{eqn:neumannmatch1}) simply indicates that $B_1(s) = 0$.  Thus (\ref{eqn:neumannmatch2}) gives 
\begin{align}
\label{eqn:Hom4}
\pdone{\phi^+_0}{n}=\pdone{\phi^-_0}{n} = 0,  \qquad \textrm{on }\Gamma,
\end{align}
so that the leading-order solution is that for a perfectly reflecting (Neumann) boundary at $\Gamma$.  
Continuing the expansion for the perforated shell, and supposing $\lambda \sim \tilde{b}_2/\eps^2 + \ldots$, the next order matching requires
\begin{align}
\label{eqn:Hom5}
\pdone{\phi_1^{\pm}}{n} = \frac{1}{2\tilde{b}_2} \big[ \phi_0 \big] \qquad \textrm{on }\Gamma.
\end{align}

\section{Shielding performance of Faraday cages\label{sec:ShieldingPerformance}}

Having derived homogenized boundary conditions for the leading order outer approximations, we now consider their shielding performance in the context of the boundary value problems introduced in \S\ref{sec:ProblemFormulation}, 
concentrating on the case when the source function $f$ is compactly supported outside of the cage, in $D \cap \Omega_+$.
For the Laplace problems the measure of good shielding is that $\nabla\phi$ should be small inside the cage interior $\Omega_-$ (since the physical field of interest is the gradient of the potential). For the Helmholtz problem we require $\phi$ itself to be small in $\Omega_-$. 

We shall illustrate our general results using explicit solutions for the special case where $\Gamma$ is the unit circle and the external forcing is due to a point source of unit strength located at a point $z_0$ outside the cage ($|z_0|>1$). Explicitly, $f=-\delta_{z_0}$, where $\delta_{z_0}$ represents a delta function supported at $z_0$. For this example we express solutions in standard polar coordinates $(\rho,\theta)$ centred at the cage centre, with $\theta=0$ corresponding to the direction of the source.
We compare the homogenized solutions with numerical solutions to the full problem in the case of disc-shaped or line-segment wires (using Model 1 to define the wire geometry).  For disc-shaped wires these are computed using the same method as \cite[Appendix A]{ChHeTr:15}; the solution is expressed as a truncated sum of radially-symmetric solutions to the Laplace or Helmholtz equation centered on the wire centers $z_j$; the coefficients in the expansion are determined by a least squares fit to the boundary conditions at discrete points on the wires.  For Laplace problems, solutions for line-segment wires can be computed using a similar method (by conformal mapping; cf. \cite{Tr:05}), although our results for this case are computed with a boundary integral equation method using \texttt{SingularIntegralEquations.jl}, a Julia package for solving singular integral equations implementing the spectral method of \cite{SIE}.

\subsection{Laplace equation with Dirichlet boundary conditions on wires}

In the case of thin wires ($\delta\ll 1$) the $\ord{1}$ outer solutions satisfy
\begin{align}
\label{eqn:outer}
 \nabla^2 \phi_0^+ = f &\qquad \textrm{in } \Omega_+, &
\nabla^2 \phi_0^- = 0 &\qquad \textrm{in } \Omega_-,\\
\label{eqn:outerbcs}
\phi_0^+ =  \phi_0^- &\qquad \textrm{on }\Gamma, &
\left[ \pdone{\phi_0}{n}\right] = \alpha\phi_0 & \qquad\textrm{on }\Gamma,
\end{align}
with $\phi_0^+$ also satisfying \rf{eqn:Infinity} at infinity. As mentioned previously, this problem is well-posed for $0<\alpha<\infty$, i.e.\ $0<\delta<\frac{\re^{-a_0}}{2\pi}$. 

\begin{figure}[t]
\centering
\includegraphics[width=\textwidth]{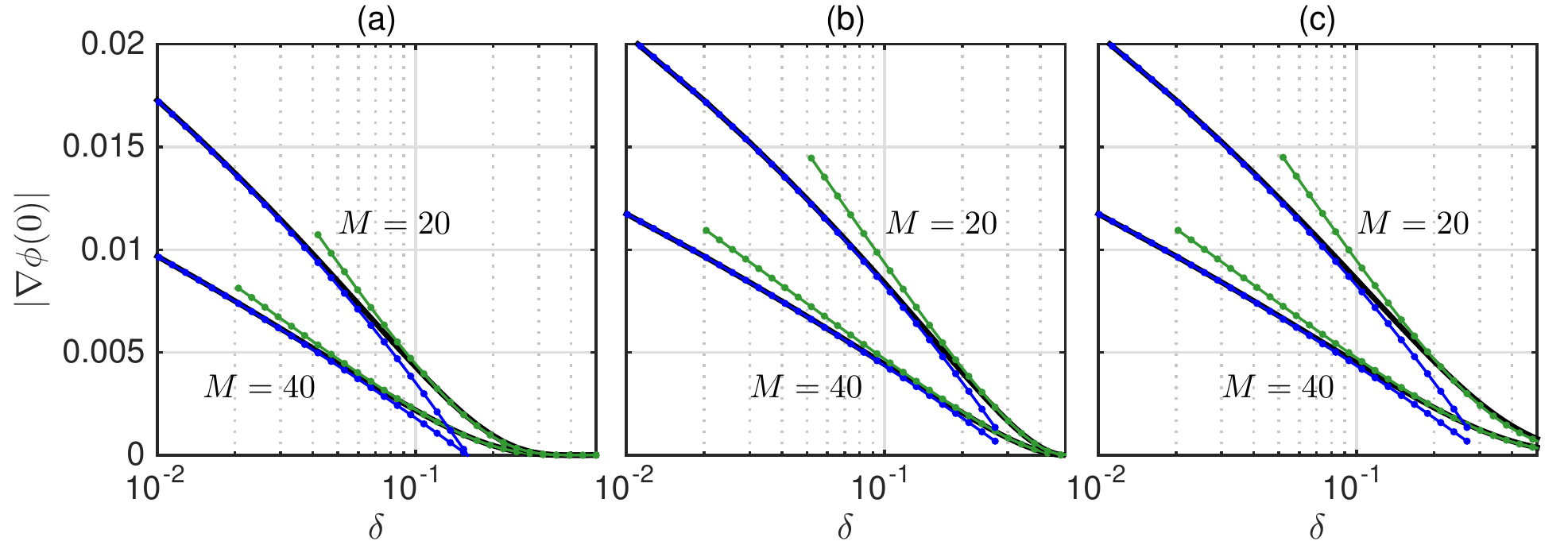}
\caption{\label{fig:electrostatic_discs}Magnitude of potential gradient at the origin for $\Gamma$ the unit circle with a source at $z_0 = 2$, for (a) circular wires, (b) perpendicular line segments, and (c) tangential line segments, for varying scaled wire radius $\delta$.  Results are shown for $\n = 20$ ($\eps = 0.314$; upper curves), and $\n = 40$ ($\eps = 0.157$).  Black lines show numerical result, blue lines/dots show the `thin-wire' asymptotic result valid for $\delta = \ord{e^{-c/\eps}}$, and green lines/dots show the `thick-wire' asymptotic result valid for $\delta = \ord{1}$.}
\end{figure}

For $\Gamma$ the unit circle and $f = -\delta_{z_0}$ the leading order solution inside the cage is
\begin{align}
\label{}
\phi^-
\sim  \phi_0^- 
= \frac{1}{\pi}\sum_{m=1}^\infty \frac{\rho^m\cos{m\theta}}{(\alpha+2m)|z_0|^m} \qquad \textrm{in }\Omega_-,
\end{align}
and in particular 
\begin{align}
\label{eqn:thinwirelaplace}
|\nabla \phi^-(0)|\sim \frac{1}{(\alpha + 2)\pi |z_0|}.
\end{align}
For shielding we need $\alpha\gg 1$, in which case $|\nabla \phi^-(0)|\sim 1/(\alpha \pi |z_0|)$. Recalling the definition of $\alpha$ in \rf{eqn:alpha}, the field inside the cage scales inverse linearly in $M$ and logarithmically in $r$, as discussed in \cite{ChHeTr:15}. 

In the case of thick wires ($\delta=\ord{1}$) 
the $\ord{1}$ outer solutions satisfy \rf{eqn:outer} but now with
\begin{align}
\label{}
\label{eqn:solidDir}
\phi_0^\pm &=  0 \qquad  \textrm{on } \Gamma,
\end{align}
with $\phi_0^+$ also satisfying \rf{eqn:Infinity} at infinity. Hence the interior and exterior problems decouple, and in particular since $\Gamma$ is a closed curve one deduces that 
\begin{align}
\label{}
\phi_0^-=0  \qquad  \textrm{in } \Omega_-.
\end{align}
Hence the leading order solution in $\Omega_-$ is the $\ord{\eps}$ term, which by \rf{eqn:Hom2} (noting that $\pdonetext{\phi_0^-}{n}=0$) satisfies the inhomogeneous Dirichlet boundary condition
\begin{align}
\label{eqn:thickouterminus}
\phi_1^-&= \tau_+\pdone{\phi^+_0}{n} \qquad  \textrm{on } \Gamma.
\end{align}
Note that only $\tau_+$ (not $\sigma_+$, $\sigma_-$ or $\tau_-$) appears in this condition for the leading order interior solution. 
The field in $\Omega_-$ is therefore $\ord{\tau_+\eps}$ as $\eps\to 0$.

For $\Gamma$ the unit circle and 
$f = -\delta_{z_0}$ the leading order solution inside the cage is
\begin{align}
\label{}
\phi^-
\sim  \eps \phi_1^- 
= \frac{\tau_+ \eps}{\pi} \sum_{m=1}^\infty \frac{\rho^m \cos{m\theta}}{|z_0|^m} \qquad \textrm{in } \Omega_-,
\end{align}
and in particular 
\begin{align}
\label{eqn:thickwirelaplace}
|\nabla \phi^-(0)|\sim \frac{|\tau_+|\eps}{\pi |z_0|}.
\end{align}

In Figure \ref{fig:electrostatic_discs} we show the excellent agreement between these approximations and the result of numerical calculations.
Note that (\ref{eqn:thinwirelaplace}) and (\ref{eqn:thickwirelaplace}) are consistent, since $\tau_+ \sim 1/\eps\alpha$ as $\delta \to 0$.

\subsection{Helmholtz equation with Dirichlet boundary conditions on wires}

In the thin wire case the analysis is similar to that for the Laplace case, with  $\phi_0^\pm$ satisfying
\begin{align}
\label{eqn:outerHelm}
(\nabla^2 + k^2) \phi_0^+ &= f \qquad  \textrm{in } \Omega_+,&
(\nabla^2 + k^2) \phi_0^- &= 0 \qquad  \textrm{in } \Omega_-,
\end{align}
the boundary conditions \rf{eqn:outerbcs}, and an outgoing radiation condition on $\phi_0^+$.

For $\Gamma$ the unit circle and 
$f = - \delta_{z_0}$ the leading order solution inside the cage is
\begin{align}
\label{eqn:CircleHelmholtzThin}
\phi^-
\sim  \phi_0^- 
&= \sum_{m=0}^\infty \frac{e_m J_m(k\rho)\cos{m\theta}}{1+\dfrac{\alpha}{k}\left(\frac{J_m'(k)}{J_m(k)}-\frac{{H_m^{(1)}}'(k)}{H_m^{(1)}(k)} \right)^{-1}} \qquad \textrm{in }\Omega_-,
\end{align}
where $e_0=(\ri/4)H_0^{(1)}(k|z_0|)$ and $e_m=(\ri/2)H_m^{(1)}(k|z_0|)$, $m\in\N$. 
In particular 
\begin{align}
\label{eqn:CircleHelmholtzThinOrigin}
\phi^-(0)\sim \frac{\frac{\ri}{4}H_0^{(1)}(k|z_0|)}{1+\dfrac{\alpha}{k}\left(\frac{{H_1^{(1)}}(k)}{H_0^{(1)}(k)}-\frac{J_1(k)}{J_0(k)} \right)^{-1}} .
\end{align}
As in the Laplace case, the field strength is $\ord{1/\alpha}$ when $\alpha\gg 1$.

In the thick wire case, at first glance the analysis appears similar to the Laplace case, with the $\ord{1}$ outer solutions satisfying \rf{eqn:outerHelm} and \rf{eqn:solidDir}. But now we must take care over the correct interpretation of \rf{eqn:solidDir}. This is because there exist resonant wavenumbers, i.e.\ values of $k$ for which $k^2$ is a Dirichlet eigenvalue of $-\nabla^2$ on $\Omega_-$, at which one cannot infer from \rf{eqn:solidDir} that $\phi_0^-$ is identically zero. We shall study such resonant cases in detail in the next section. Here we simply record that, if we ignore resonance effects and assert that $\phi_0^-=0$, the leading order solution in $\Omega_-$ is again provided by the $\ord{\eps}$ term, which satisfies \rf{eqn:thickouterminus}, just as in the Laplace case.

For $\Gamma$ the unit circle and 
$f = - \delta_{z_0}$ the leading order non-resonant solution inside the cage is
\begin{align}
\label{eqn:CircleHelmholtzThick}
\phi^-
\sim  \eps \phi_1^- 
&= k\eps \tau_+ \sum_{m=0}^\infty e_m \left(\frac{J_m'(k)}{J_m(k)}-\frac{{H_m^{(1)}}'(k)}{H_m^{(1)}(k)} \right)J_m(k\rho)\cos{m\theta} \qquad \textrm{in }\Omega_-,
\end{align}
where $e_m$ are as above. 
In particular 
\begin{align}
\label{eqn:CircleHelmholtzThickOrigin}
\phi^-(0)\sim k\eps \tau_+ \frac{\ri}{4}H_0^{(1)}(k|z_0|) \left(\frac{{H_1^{(1)}}(k)}{H_0^{(1)}(k)} - \frac{J_1(k)}{J_0(k)} \right).
\end{align}

\begin{figure}[t]
\centering
\includegraphics[width=\textwidth]{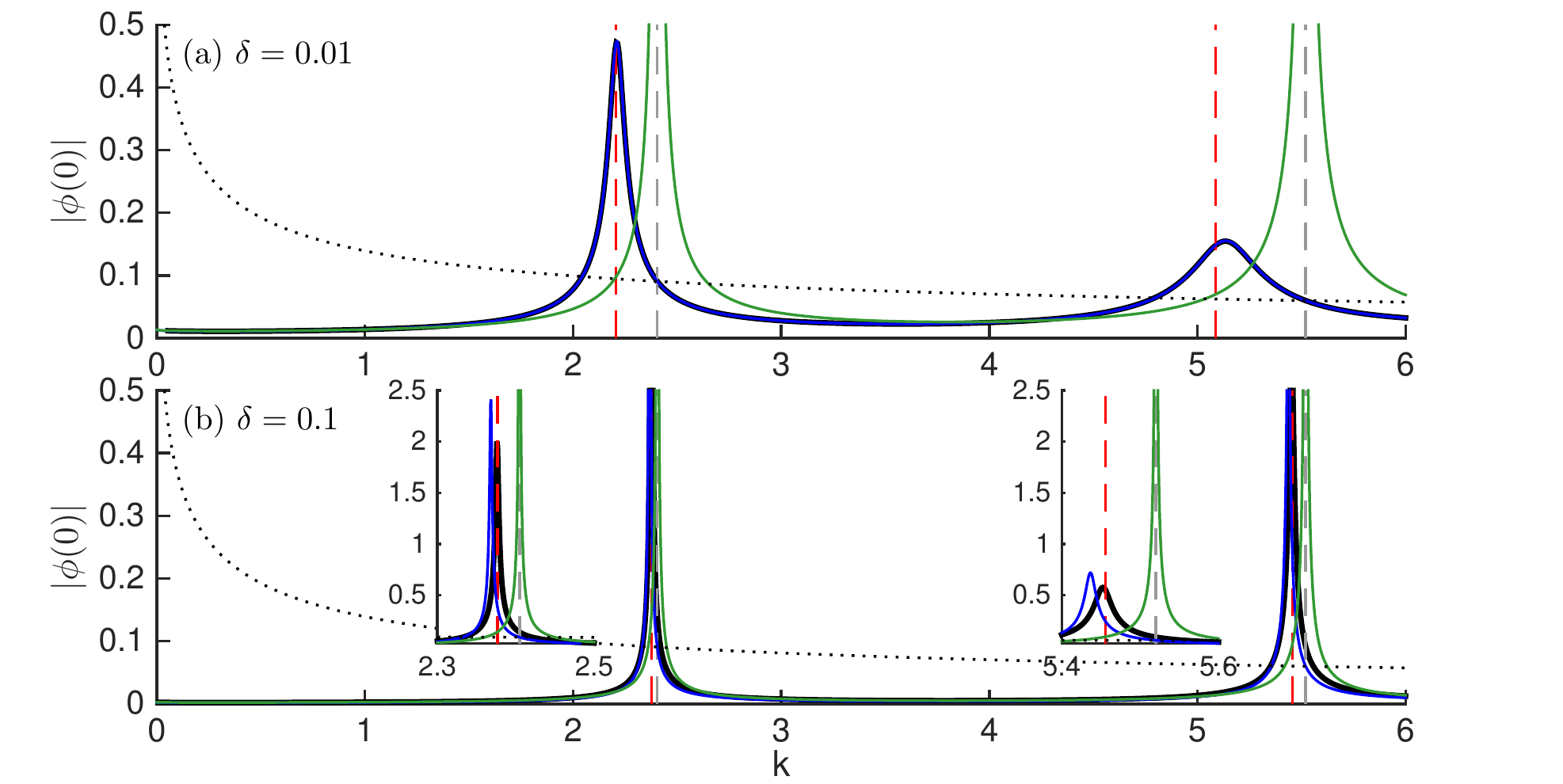}
\caption{\label{fig:150930ae}Amplitudes at $z = 0$ for the wave problem for disk-shaped wires arranged around the unit circle, for varying wavenumber $k$. Corresponding field plots for particular wavenumbers are shown in Figure \ref{fig:151003a}. Parameters are $\n = 30$, $z_0 = 2$, and (a) $\delta = 0.01$, (b) $\delta = 0.1$.  Black lines show the numerical solution, blue lines show the `thin-wire' asymptotic result, green lines shows the `thick-wire' asymptotic result (without correcting for resonance), and the dotted line shows the unshielded (free-field) solution.    Vertical grey dashed lines indicate the unperturbed resonances for the unit circle corresponding to axisymmetric modes (two asymmetric modes are also excited in this wavenumber range, but have zero amplitude at the origin), and red dashed lines show the shifted resonances calculated using the $\ord{\eps}$ perturbation in \S\ref{sec:ShieldingPerformance}\ref{sec:resonance}.   Insets in the lower panel show enlargements around the peaks.}
\end{figure}

\begin{figure}[t]
\centering
\includegraphics[width=\textwidth]{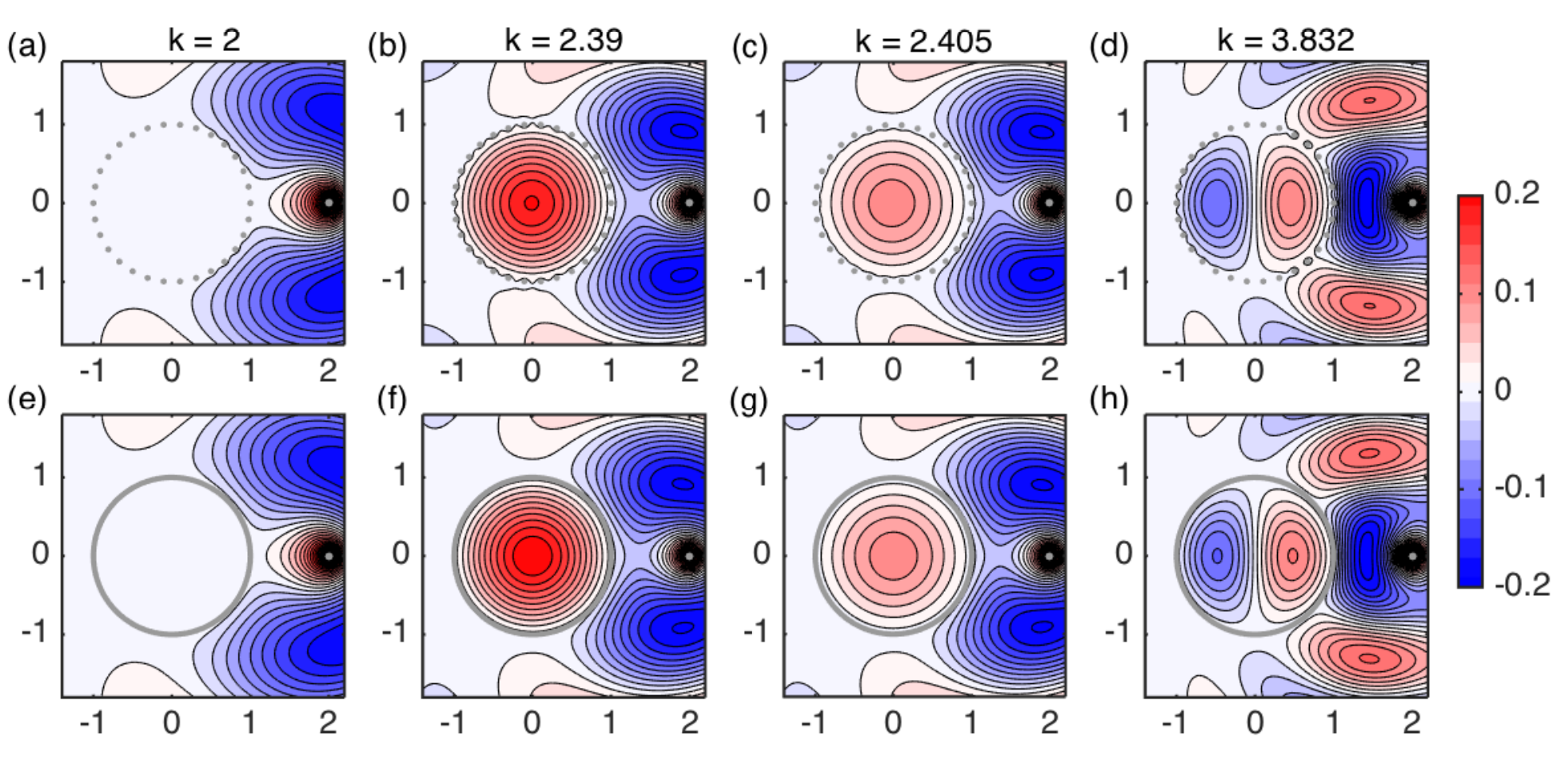}
\caption{\label{fig:151003a}(a)-(d) Numerical solutions to the wave problem for disk-shaped wires arranged around the unit circle, for four different wavenumbers $k$, showing $\Re(\phi(z))$. Parameters are $\n = 30$, $\delta = 0.1$, and $z_0 = 2$. (b)-(d) represent near-resonant cases: in (b) and (c) the relevant resonant mode is $J_{0}(k|z|)$, $k\approx 2.405$, and in (d) it is $J_{1}(k|z|) \cos(\arg(z))$, $k\approx 3.382$. (e)-(h) `Thick-wire' asymptotic solutions for the same problems; in (e) this is the non-resonant solution \rf{eqn:CircleHelmholtzThick}, and in (f)-(h) we plot the leading order resonance-corrected solution from \S\ref{sec:ShieldingPerformance}\ref{sec:resonance}, i.e. $\phi_0^+$ in $\Omega_+$ and $\phi_{-1}^-$ in $\Omega_-$.}
\end{figure}

When one compares the approximations \rf{eqn:CircleHelmholtzThinOrigin} and \rf{eqn:CircleHelmholtzThickOrigin} with numerical simulations for fixed $k$ away from resonance, one observes similar behaviour to that in Figure \ref{fig:electrostatic_discs}, i.e.\ \rf{eqn:CircleHelmholtzThinOrigin} is accurate for small $\delta$ and \rf{eqn:CircleHelmholtzThickOrigin} for larger $\delta$. However, interesting new behaviour become apparent when one fixes $\delta$ and varies the wavenumber $k$. Two plots of this type are presented in Figure \ref{fig:150930ae}. One finds that close to resonant wavenumbers the numerical solution is strongly peaked, and the amplitude $|\phi(0)|$ can actually exceed that of the free-field solution; that is, the cage \emph{amplifies} the field rather than shielding from it. This amplification is clear in the near-resonant field plots in Figure \ref{fig:151003a}.

Returning to Figure \ref{fig:150930ae}, we note that the position of the peak amplitude is in general slightly shifted from the exact resonance. For sufficiently small $\delta$ (cf.\ Figure \ref{fig:150930ae}(a)) the peaks are captured correctly by the `thin-wire' asymptotic result. But for larger $\delta$ the position and height of the peak are not predicted correctly (cf.\ Figure \ref{fig:150930ae}(b)). Unfortunately the `thick wire' approximation \rf{eqn:CircleHelmholtzThick} cannot capture the peaks either - the $\ord{\eps}$ term $\phi^{-}_1$ blows up to infinity at the exact resonances, as is obvious from \rf{eqn:CircleHelmholtzThick}, and our asymptotic solution breaks down. 
In the next section we show how the `thick wire' approximation \rf{eqn:CircleHelmholtzThickOrigin} can be modified to correctly predict the near-resonant behaviour for larger values of $\delta$.

\subsection{Resonance effects\label{sec:resonance}}

Close to resonant wavenumbers, our thick wire ($\delta = \ord{1}$) solution \rf{eqn:CircleHelmholtzThick} breaks down, as the assertion that $\phi_0^-=0$ is invalid. Instead, we expect a near-resonant response in which the leading order interior solution is a non-trivial linear combination of the corresponding eigenmodes. 

To examine the behaviour close to resonance, let $k = k_* + \eps \tilde{k}$, where $k_*>0$ is a resonant wavenumber with real-valued eigenmode $\psi$ satisfying $(\nabla^2 + k_*^2) \psi = 0$ in $\Omega_-$ and $\psi=0$ on $\Gamma$, and $\tilde{k}=\ord{1}$. (For simplicity we shall always assume there is only one eigenmode corresponding to $k_*$; more generally we would have a superposition of eigenmodes). Expanding \rf{eqn:Helmholtz} with $\phi=\phi_0^\pm+\eps \phi_1^\pm + \ord{\eps^2}$ as in \rf{eqn:2t0}, the leading order interior solution satisfies
\begin{align}
\label{eqn:LeadingOrderInsideres}
(\nabla^2 + k_*^2)  \phi^{-}_{0} = 0& \qquad \textrm{in } \Omega_-, \\
\phi^{-}_{0} = 0& \qquad \textrm{on } \Gamma,
\label{eqn:LeadingOrderInsideres3}
\end{align}
whence
\begin{align}
\label{eqn:LeadingOrderInsideres2}
\phi^{-}_{0} = C_{0} \psi,
\end{align}
for some amplitude $C_{0}$ to be determined. By \rf{eqn:Hom2} the next order interior problem is
\begin{align}
\label{eqn:FirstOrderInsideres}
(\nabla^2 + k_*^2)  \phi^{-}_{1} = -2k_* \tilde{k} C_{0} \psi & \qquad \textrm{in } \Omega_-, \\
\phi^{-}_{1} = \tau_+ \pdone{{\phi}^{+}_0}{n} -\sigma_- C_{0} \pdone{\psi}{n} & \qquad \textrm{on } \Gamma,
\end{align}
where the inhomogeneous term on the right hand side of \rf{eqn:FirstOrderInsideres} arises from the perturbation of the eigenvalue from $k_*$. Since the associated homogeneous problem has a non-zero solution, $\psi$, there is a solvability condition to be satisfied, following from the identity
\begin{align}
\label{eqn:SolvabilityIdentity}
\int_{\Omega_-} \psi \left( (\nabla^2 + k_*^2)  \phi^{-}_{1} \right) \,{\rm d}S = 
- \int_{\Gamma}  \phi^{-}_1 \pdone{\psi}{n} \,{\rm d}s,
\end{align}
which can be obtained using Green's second identity.
Defining 
\begin{align}
\label{eqn:integrals1to3}
 I_1 = \int_{\Omega_-} \psi^2 \,{\rm d}S, \qquad 
 I_2 = \int_{\Gamma} \left(\pdone{\psi}{n}\right)^2 \,{\rm d}s, \qquad
 I_3 =  \int_{\Gamma} \kappa \left(\pdone{\psi}{n}\right)^2 \,{\rm d}s,
\end{align}
($I_3$ is for later use), the solvability condition arising from \rf{eqn:SolvabilityIdentity} is that
\begin{align}
\label{eqn:Solvability}
(2k_* I_1 \tilde{k}  + \sigma_- I_2) C_{0} = \tau_+ \int_{\Gamma} \pdone{{\phi}^{+}_0}{n}\pdone{\psi}{n} \,{\rm d}s.
\end{align}
This determines the amplitude $C_0$ of the $\ord{1}$ interior solution \rf{eqn:LeadingOrderInsideres2}, 
except when %
\begin{align}
\label{eqn:Shiftres}
\tilde{k} = \tilde{k}_* := -\frac{\sigma_-I_2}{2k_*I_1},
\end{align}
where $C_0$ blows up to infinity. This represents a shift in the position of the apparent resonance from the original value $k_*$ to the perturbed value $k_* + \eps \tilde{k}_*$. We note that the shift $\tilde{k}_*$ depends both on the wire shape $K$ (through $\sigma_-$) and on the cage geometry $\Gamma$ (through $I_1$ and $I_2$). 
Furthermore, we note that the sign of the shift is given by the sign of $-\sigma_-$. For line segment wires parallel to $\Gamma$, $\sigma_-$ is positive for all $0<\delta<1/2$, so the shift is always negative. But in general there may exist a critical value of $\delta$ at which $\sigma_-$ (and hence the shift) changes sign. For circular wires this occurs at $\delta\approx 0.12$ (cf.\ Figure \ref{fig:inner1}). 

The true solution is not actually infinite at the shifted value $k=k_* + \eps \tilde{k}_*$; rather there is a narrow region of $\ord{\eps^2}$ around this value in which the amplitude of the interior solution is large. %
To capture this behaviour 
we write $k = k_* + \eps \tilde{k}_* + \eps^2 \tilde{\tilde{k}}$, where $\tilde{k}_*$ is as in (\ref{eqn:Shiftres}) and 
$\tilde{\tilde{k}}=\ord{1}$, and  %
introduce an extra leading term in the expansion of the interior solution,
\begin{align}
\label{eqn:InsideRes}
\phi^-(x,y)= \frac{1}{\eps} \phi^-_{-1}(x,y) + \phi^-_0(x,y) + \eps\phi^-_1(x,y) + \ord{\eps^2} \qquad \textrm{in } \Omega_-.
\end{align}

As a result we require an additional $\ord{1}$ term in the boundary layer solution, which becomes
\begin{align}
\label{eqn:InsideRes2}
\Phi(N,S,s)=  & -\pdone{\phi_{-1}^{-}}{n}(s) \Phi^-(N,S) - \eps \kappa(s) \pdone{\phi_{-1}^{-}}{n}(s) \tilde{\Phi}^{-}(N,S) - \eps \frac{\partial^2 \phi_{-1}^{-}}{\partial n\partial s}(s) \hat{\Phi}^{-}(N,S)  \nonumber\\ 
& \quad +\eps \pdone{\phi^+_{0}}{n}(s)\Phi^+(N,S) - \eps \pdone{\phi^-_{0}}{n}(s)\Phi^-(N,S) + \ord{\eps^2},
\end{align}
where the functions $\hat{\Phi}^\pm$ and $\tilde{\Phi}^\pm$ are defined in Appendix \ref{sec:higherorder}. 
This solution is obtained from the general solution to the boundary layer problem given in Appendix \ref{sec:higherorder}, choosing the constants in that solution to match the gradients of the interior and exterior outer expansions. If $\phi_{-1}^-= 0$ it reduces to the solution given earlier.
The resulting matching conditions for the outer solutions, analogous to (\ref{eqn:MatchPlus2})-(\ref{eqn:MatchMinus2}), are
\begin{align}
\label{eqn:MatchPlus2Res}
- (\tau_- - \eps \kappa \tilde{\tau}_-) \pdone{\phi^-_{-1}}{n} - \eps \hat{\tau}_- \frac{\partial^2 \phi_{-1}^{-}}{\partial n\partial s} + \eps\sigma_+\pdone{\phi^+_0}{n} - \eps\tau_-\pdone{\phi^-_0}{n} = \phi_0^+ + \eps\phi^+_1 &\qquad \textrm{on }\Gamma,\\
\label{eqn:MatchMinus2Res}
-(\sigma_- - \eps \kappa \tilde{\sigma}_-)\pdone{\phi^-_{-1}}{n} - \eps \hat{\sigma}_- \frac{\partial^2 \phi_{-1}^{-}}{\partial n\partial s} +
 \eps\tau_+\pdone{\phi^+_0}{n} -\eps\sigma_-\pdone{\phi^-_0}{n} = \frac{1}{\eps} \phi_{-1}^- + \phi_0^- + \eps \phi^-_1  &\qquad \textrm{on }\Gamma,
\end{align}
where $\hat{\sigma}_{\pm}$, $\hat{\tau}_{\pm}$, $\tilde{\sigma}_{\pm}$ and $\tilde{\tau}_{\pm}$ are far field constants in the expansions of $\hat{\Phi}^\pm$ and $\tilde{\Phi}^\pm$ (these constants may depend on the choice of wire model; see Appendix \ref{sec:higherorder}).

The leading order interior problem for $\phi_{-1}^-$ is identical to the earlier problem  (\ref{eqn:LeadingOrderInsideres})-(\ref{eqn:LeadingOrderInsideres3}), with solution
\begin{align}
\label{eqn:LeadingOrderInsideRes2}
\phi^{-}_{-1} = C_{-1} \psi,
\end{align}
where $C_{-1}$ is to be determined. 
This large interior solution causes a change to the leading order exterior problem, for which the boundary condition (from (\ref{eqn:MatchPlus2Res})) becomes
\begin{align}
\label{eqn:LeadingOrderOutsideRes}
\phi^{+}_{0} = - \tau_- C_{-1} \pdone{\psi}{n} \qquad \textrm{on } \Gamma.
\end{align}
We split 
$\phi_0^+$ into two components: 
one due to the source, and one forced by 
the boundary condition \rf{eqn:LeadingOrderOutsideRes}, writing
\begin{align}
\label{eqn:LeadingOrderOutsideRes2}
\phi^{+}_{0} = \hat{\phi}^{+}_0 + \tau_- C_{-1} \tilde{\phi}^{+}_0,
\end{align}
where $(\nabla^2 + k_*^2)  \hat{\phi}^{+}_{0} = f$ in $\Omega_+$ with $\hat{\phi}_0^+=0$ on $\Gamma$ and 
$(\nabla^2 + k_*^2)  \tilde{\phi}^{+}_{0} = 0$ in $\Omega_+$ with $\tilde{\phi}_0^+=-\pdonetext{\psi}{n}$ on $\Gamma$, with both $\hat{\phi}_0^+$ and $\tilde{\phi}_0^+$ satisfying outgoing radiation conditions at infinity.

The $\ord{1}$ interior problem is
\begin{align}
\label{eqn:FirstOrderInsideRes1}
(\nabla^2 + k_*^2)  \phi^{-}_{0} = -2k_* \tilde{k}_* C_{-1} \psi & \qquad \textrm{in } \Omega_-, \\
\label{eqn:FirstOrderInsideRes2}
\phi^{-}_{0} =  -\sigma_- C_{-1} \pdone{\psi}{n} & \qquad \textrm{on } \Gamma.
\end{align}
The solvability condition is the same as (\ref{eqn:Solvability}) but with zero right-hand side and $C_0$ replaced with $C_{-1}$.  This holds identically, given the definition of $\tilde{k}_*$ (cf.\ (\ref{eqn:Shiftres})), 
so the amplitude $C_{-1}$ remains undetermined at this order.  
Writing the solution to \rf{eqn:FirstOrderInsideRes1}-\rf{eqn:FirstOrderInsideRes2} as
\begin{align}
\phi^{-}_0 = \sigma_- C_{-1} \tilde{\phi}^{-}_0 + C_0 \psi, 
\end{align}
where $\tilde{\phi}^{-}_{0}$ is a particular solution of $(\nabla^2 + k_*^2)  \tilde{\phi}^{-}_{0} = (I_2/I_1)\psi$ in $\Omega_-$ with $\tilde{\phi}_0^-=-\pdonetext{\psi}{n}$ on $\Gamma$, and $C_0$ is arbitrary,
the $\ord{\eps}$ interior problem becomes
\begin{align}
\label{eqn:SecondOrderInsideRes}
(\nabla^2 + k_*^2) \phi^{-}_{1} = -2k_* \tilde{k}_* C_0 \psi - C_{-1}\left(2k_* \tilde{k}_* \sigma_-  \tilde{\phi}^{-}_0  + (\tilde{k}_*^2 + 2k_* \tilde{\tilde{k}})\psi \right) &\qquad \textrm{in } \Omega_-, \\ 
 \phi^{-}_{1} =  \tau_+ \pdone{\hat{\phi}^{+}_0}{n} - \sigma_-C_0 \pdone{\psi}{n}  
+ C_{-1}\left(\tau_+ \tau_-  \pdone{\tilde{\phi}^{+}_0}{n} -\sigma_-^2  \pdone{\tilde{\phi}^{-}_0}{n} +\kappa \tilde{\sigma}_-\pdone{\psi}{n} - \hat{\sigma}_-  \frac{\partial^2 \psi}{\partial n\partial s}\right) & \qquad \textrm{on } \Gamma.
\end{align}
Note that the right hand sides now contains terms due to the exterior field, as well as lower-order components of the interior field. The solvability condition is %
\begin{align}
\label{eqn:Solvability3}
\left(\left(\tilde{k}_*^2  + 2 k_* \tilde{\tilde{k}}\right)I_1 - \tilde{\sigma}_- I_3 - \tau_+ \tau_- I_4 + \sigma_-^2 I_5  + 2 k_* \tilde{k}_* \sigma_- I_6  \right) C_{-1} = \tau_+ I_7,
\end{align}
where
\begin{align}
\label{eqn:integrals4to7}
&
I_4 = \int_{\Gamma} \pdone{\tilde{\phi}^{+}_0}{n}\pdone{\psi}{n} \,{\rm d}s, \quad
I_5 = \int_{\Gamma} \pdone{\tilde{\phi}^{-}_0}{n}\pdone{\psi}{n}\,{\rm d}s, \quad
I_6 = \int_{\Omega_-} \tilde{\phi}^{-}_0 \psi \,{\rm d}S, \quad
I_7 = \int_{\Gamma} \pdone{\hat{\phi}^{+}_0}{n}\pdone{\psi}{n} \,{\rm d}s.  \quad 
\end{align}
In deriving \rf{eqn:Solvability3} from \rf{eqn:SecondOrderInsideRes} the $C_0$ terms cancel due to (\ref{eqn:Shiftres}), and the term proportional to $\hat{\sigma}_-$ integrates to zero since $\Gamma$ is a closed loop.
Noting that $I_4$ and $I_7$ are in general complex, whereas $I_1$, $I_3$, $I_5$ and $I_6$ are real, the condition \rf{eqn:Solvability3} determines $C_{-1}$ with
\begin{align}
\label{eqn:CMinus1Formula}
|C_{-1}| =  |A| a \left(( \tilde{\tilde{k}} - \tilde{\tilde{k}}_*)^2 +a^2\right)^{-1/2},
\end{align}
where
\begin{align}
\label{eqn:Ampres}
A = \frac{I_7}{\tau_- \Im(I_4)}, \quad
a = \frac{\tau_+\tau_- \Im(I_4)}{2I_1 k_*}, \quad
\tilde{\tilde{k}}_* = - \frac{ I_1 \tilde{k}_*^2  - \tilde{\sigma}_- I_3 - \tau_+\tau_- \Re(I_4) + \sigma_-^2 I_5 + 2 k_* \tilde{k}_* \sigma_- I_6 }{ 2I_1 k_*}.
\end{align}
From \rf{eqn:CMinus1Formula} it follows that the maximum of $|C_{-1}|$ is 
 $|A|$ at $\tilde{\tilde{k}} = \tilde{\tilde{k}}_*$. So, to conclude, the near-resonant response occurs in a range of wavenumbers of width $\ord{ \tau_+\tau_- \eps^2}$ around $k = k_* + \eps \tilde{k}_* + \eps^2 \tilde{\tilde{k}}_*$, and the maximum amplitude is $\ord{1/(\tau_- \eps )}$.  The exterior field remains $\ord{1}$.

\begin{figure}[t]
\centering
\includegraphics[width=\textwidth]{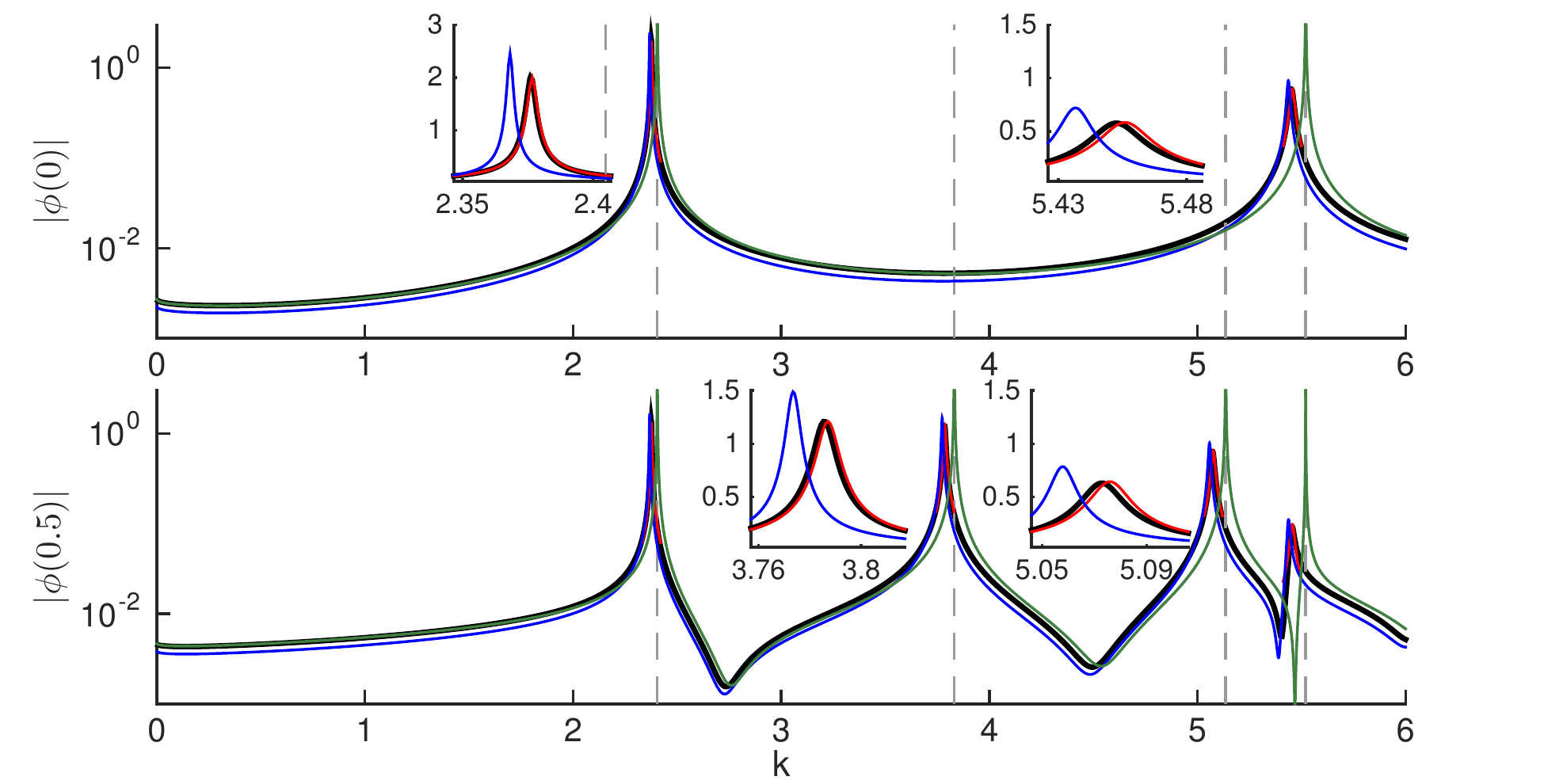}  %
\caption{\label{fig:150930a}Amplitudes at $z = 0$ and $z = 0.5$ for the wave problem for disk-shaped wires arranged around the unit circle, for varying wavenumber $k$. Parameters are $\n = 30$, $\delta = 0.1$, and $z_0 = 2$.  Insets show enlargements of the regions close to resonance.  Black line shows the numerical solution, blue line shows the `thin-wire' asymptotic result, green line shows the non-resonant `thick-wire' asymptotic result, and red line shows the resonant thick-wire result.  Vertical dashed lines indicate the unperturbed resonances for the unit circle.}
\end{figure}

\begin{figure}[t]
\centering
\includegraphics[width=\textwidth]{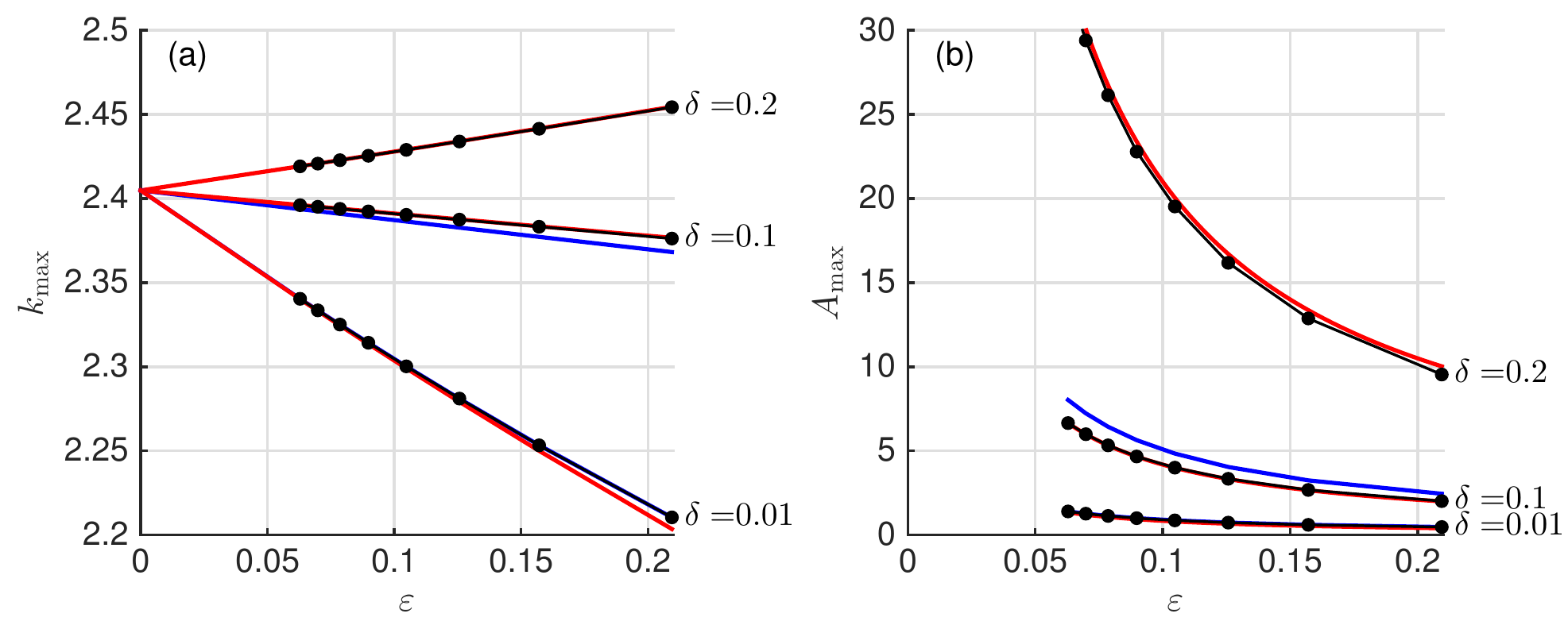}
\caption{\label{fig:150924a} (a) Wavenumbers giving maximum amplitude close to the first resonance for disk-shaped wires arranged around the unit circle, for varying $\eps$ (number of wires) and for different scaled wire radius $\delta$, together with (b) the amplitude (at $z = 0$) for that wavenumber.  Black lines/dots show maxima from the numerical solutions, blue lines show the maxima from the `thin-wire' asymptotic solution, and red lines show the `thick-wire' asymptotic result.  The thin-wire result is not shown for $\delta = 0.2$ since $\alpha<0$ in that case, so that approximation is invalid; for $\delta = 0.01$, the thin wire result is indistinguishable from the numerical solution in this plot.}
\end{figure}

The good agreement between these predictions and the result of numerical calculations is shown in Figures \ref{fig:151003a}, \ref{fig:150930a} and \ref{fig:150924a}.   The insets in figure \ref{fig:150930a} demonstrates that the shape of the amplitude variation with wavenumber near the resonance is well captured, and Figure \ref{fig:150924a} demonstrates how the position and amplitude at the peak vary with $\eps$.  We emphasise that as the number of wires increases, the resonant response occurs closer to the unperturbed resonant modes of $\Omega_-$, over an increasingly narrow band of wavenumbers, but with an increasingly large amplitude.

\subsection{Neumann solutions and resonance effects}

For the equivalent problems satisfying Neumann conditions on the wires, we have seen in \S\ref{sec:EquivBCs} that there is in general much weaker shielding than for Dirichlet conditions.  Unless the gaps between the wires are small, the leading order homogenized solution does not notice the wires at all, and even for small gaps the homogenized wires provide a jump condition on $\Gamma$ that does not necessarily lead to a weak field inside the cage.  Only in the case of `very small gaps' is there a significant shielding effect.   Although this is not the main focus of our study (requiring very small gaps largely defeats the idea of a Faraday cage), we touch briefly on this very small gap case because of its analogy to the Dirichlet problems above.  In particular, we focus on the perforated shell introduced in \S\ref{sec:EquivBCs}, for which the homogenized boundary conditions are (\ref{eqn:Hom4}) and (\ref{eqn:Hom5}),
which depend on $ \tilde{b}_2 \sim \eps^2 \lambda$ as determined from the solution to the boundary-layer cell problem.

For the Laplace problem, the $\ord{1}$ solutions satisfy $\nabla^2 \phi_0^- = 0$ subject to the homogenous Neumann conditions (\ref{eqn:Hom4}) on $\Gamma$.  
The interior solution $\phi_0^-$ is therefore a constant, which is determined from the solvability condition on the next-order problem: $\nabla^2 \phi_1^- = 0$ with (\ref{eqn:Hom5}) on $\Gamma$.
This determines the constant $\phi_0^-$ to be the average of the exterior solution, $\phi_0^+$, around $\Gamma$.  The correction, which controls the size of $|\nabla \phi(0)|$, is $\ord{1/(\eps\lambda)}$. 

For the Helmholtz problem, the $\ord{1}$ solutions satisfy (\ref{eqn:outerHelm}) subject to homogenous Neumann conditions (\ref{eqn:Hom1}) on $\Gamma$.  Away from resonance, the interior solution is $\phi_0^- = 0$, and the correction is again $\ord{1/(\eps \lambda)}$.  As for the Dirichlet problem, however, this solution breaks down if $k$ is close to a resonant wavenumber $k_*$ for which there is a non-zero solution $\psi$ to $(\nabla^2 + k_*^2 )\psi = 0$ in $\Omega_-$ with $\partial \psi/\partial n = 0$ on $\Gamma$. 
The resonant case can be analysed in an equivalent fashion to the Dirichlet problem.  Without giving the details, we find that the wavenumber is shifted to $k = k_* + (1/\eps\lambda)\, I_2/4k_* I_1 + \ord{1/(\eps\lambda)^2}$, where $I_1$ and $I_2$ are as defined in (\ref{eqn:integrals1to3}) for the relevant eigenfunction, while the peak amplitude at the origin is $\ord{\eps\lambda}$.  Recall that in terms of the scaled gap size $1/2-\delta$, we have $\eps\lambda \sim (\eps/\pi) \log(1/(\pi(\half-\delta)))$, so this resonant amplitude grows logarithmically as the size of the gaps is reduced.  
\section{Discussion and Conclusions\label{sec:Discussion}}

We have derived homogenized boundary conditions for various instances of the two-dimensional Faraday cage problem, helping to quantify the effect of a wire mesh on electrostatic and electromagnetic shielding. We have given an overview in \S\ref{sec:EquivBCs} of the different leading order behaviour that can occur depending on the scaled wire size $\delta$, extending previous results for the `thin wire' regime $\delta \ll 1$, and incorporating the effects of finite wire size that in general allow for better shielding.  The homogenized conditions help to clarify how the wire geometry affects the shielding behaviour, through the solution of cell problems and extraction of far-field constants.  This allows us to make some general comments on the shielding efficiency of different wires. 
For brevity we focus mainly on the case of Dirichlet boundary conditions.

In the Dirichlet case, we showed that when the exterior wave field is $\ord{1}$, the interior field is generally $\ord{ \tau_+ \eps}$, where $\eps = |\Gamma|/M$ and $\tau_+$ encodes the wire geometry.  For thin wires we established the approximation (\ref{eqn:MuNuDeltaSmall}) for $\tau_+$, which indicates that the logarithmic capacity of the wires (controlled by their size and shape) is the key property governing shielding. For thicker wires, the orientation of the wires also becomes important, and the parameter $\tau_+$ can become small when the gap between wires is small. In this regime the relationship between the gap thickness (expressed as a fraction of the length of $\Gamma$) and the size of $\tau_+$ is strongly dependent on the wire shape. 
For example $\tau_+ = 0.01$ is achieved with a gap thickness of approximately $0.22$ for tangential line segments, but as much as $0.54$ for circular wires, and $0.61$ for square wires. (For perpendicular line segments, the gap thickness is always 1, but a wire length of $2\delta \approx 1.12$ is required to achieve a correspondingly small value of $\tau_+$).

We also derived a model for resonance effects in Faraday cages, showing how the incident exterior wave field can be amplified by the presence of the cage in a narrow range of wavenumbers close to the resonant wavenumbers for the corresponding solid shell.  The analysis showed that at its peak this resonance gives rise to a wave field $\ord{1/(\tau_-\eps)}$ larger than the incident field, and that this occurs over a range of wavenumbers of width $\ord{\tau_-\tau_+\eps^2}$.

A similar analysis applies for a source inside the cage, when it is desired to shield the exterior region (as for a microwave oven, for example).  In that case, for the `thick-wire' regime, away from resonance the interior solution is $\ord{1}$ and the exterior field is $\ord{\tau_- \eps}$.  Resonance occurs at the same shifted eigenvalues as for the exterior source problem, but the peak amplitude is now $\ord{1/(\tau_-\tau_+\eps^{2})}$, and the corresponding radiated field outside the cage is $\ord{1/(\tau_+\eps)}$.  (The relative change in field strength from the non-resonant case is the same as in the case of an exterior source).  Essentially the same analysis as in \S\ref{sec:ShieldingPerformance}\ref{sec:resonance} can be followed, with the same result except that (\ref{eqn:CMinus1Formula}) gives the amplitude of the $\ord{1/\eps^2}$ interior solution, and the forcing term $\tau_+ I_7$ in (\ref{eqn:Ampres}) is replaced with
\begin{align}
\label{eqn:I8formula}
I_8 = \int_{\Omega_-} f \psi \,{\rm d}S.
\end{align}

\begin{figure}[t]
\centering
\includegraphics[width=\textwidth]{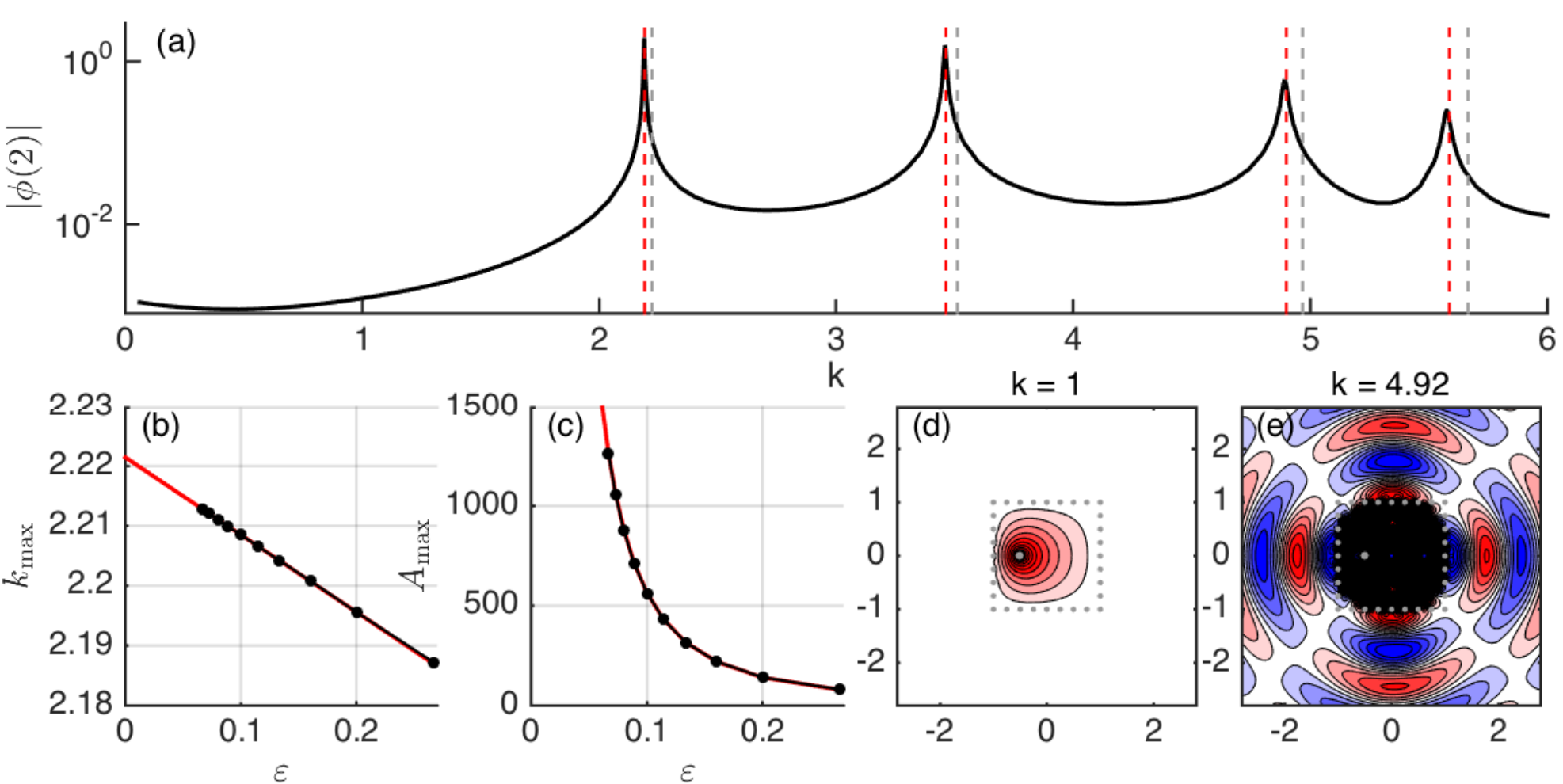}
\caption{\label{fig:160102abc}Solutions for the wave problem for disk-shaped wires arranged around the unit square with source at $z_0 = -0.5$.  (a) Numerically calculated amplitudes of the wave field at $z = 2$ for varying wavenumber $k$, with parameters $\n = 32$ ($\eps = 0.125$), $\delta = 0.1$.   Vertical grey dashed lines indicate the unperturbed resonances for the unit square, and red dashed lines show the shifted resonances calculated using the $\ord{\eps}$ perturbation from (\ref{eqn:Shiftres}).  (b) Wavenumber giving maximum amplitude close to the first resonance for varying $\eps$, together with (c) the maximum amplitude (at $z = 0$) for that wavenumber. Black lines/dots show maxima from the numerical solutions, red lines show the asymptotic solutions for $k_*+\eps\tilde{k}_*$ from (\ref{eqn:Shiftres}) and for $I_8/(\tau_-\tau_+\eps^2 \Im(I_4))$ from (\ref{eqn:Ampres}) with the modification in (\ref{eqn:I8formula}).  (d)-(e) Example numerical solutions showing $\Re(\phi(z))$, away from resonance, and close to one of the resonant wavenumbers.}
\end{figure}

Although our homogenized boundary conditions were derived for smooth cages $\Gamma$,  applying the resulting models to non-smooth geometries appears to give reasonable results, at least in terms of computing resonance shifts and amplitudes.  As an example of both this, and the interior source, we consider a cage of circular wires arranged on a unit square, with a point source located inside the cage at $z = -0.5$.  Numerical solutions illustrating the resonance effects are shown in Figure \ref{fig:160102abc}. 

The unperturbed resonances for this problem are $k_* = (\pi/2) (l^2+m^2)^{1/2}$, $l,n\in\N$, for which 
\begin{align}
I_1 = 1, \ \qquad 
I_2 =  \frac{\pi}{2} (l^2+m^2)^{1/2}.
\end{align}
To calculate amplitude and corrections, we need to solve for $I_4$ and $I_8$.  For the first resonance ($l = m = 1$), numerical solution of the relevant exterior problem for $\tilde{\phi}_0^+$ (performed using the \texttt{MPSpack} software package, which implements the non-polynomial finite element method of \cite{BaBe:10}) gives $I_4 \approx 3.00 - 16.02i$, while $I_8 = 1/\sqrt{2}$.
As Figure \ref{fig:160102abc} shows, the analysis appears to capture the $\ord{\eps}$ resonance shift correctly, as well as the $\ord{1/\eps^2}$ variation of the peak amplitude.  To gain more accuracy in the resonance shift we expect it would be necessary to consider local approximations in the vicinity of the corners (which were neglected in our anlaysis) and match these to the boundary layer and outer expansions, following the procedure outlined in \cite{Na:08a,Na:08b,DeScSe:15}.

Our analysis of the Neumann problem shows that, as one might expect, Neumann wires shield much less effectively than Dirichlet wires of the same size and shape. For the acoustic problem this implies that it is very difficult to shield noise using a mesh-like structure made of sound-hard material unless the gaps are very small. The implication for the electromagnetic problem is that a cage of parallel wires may provide reasonable shielding of waves whose electric field is polarized parallel to the wire axes, but will not shield waves whose electric field is polarized perpendicular to the wires axes. This effect is the basis of many polarizing filters, and explains, at least intuitively, why the mesh in the doors of microwave ovens is made of a criss-cross wire pattern or a perforated sheet, rather than from parallel wires aligned in a single direction. However, a proper treatment of the full 3D electromagnetic case is left for future work.

\section{Funding}
IJH is supported by a Marie Curie FP7 Career Integration Grant within the 7th European Union Framework Programme. 
\section{Acknowledgements}
The authors gratefully acknowledge helpful discussions with Nick Trefethen, Jon Chapman, Mikael Slevinsky and Alex Barnett. 
\bibliography{Faraday_bib}
\bibliographystyle{siam}
\appendix

\section{Higher order boundary layer expansions\label{sec:higherorder}}

In this section we outline the derivation of the two-term boundary layer expansion 
\begin{align}
\label{eqn:BLTwoTerm}
\Phi(N,S;s) = \Phi_0(N,S;s) + \eps \Phi_1(N,S;s) + \ord{\eps^2},
\end{align}
extending the leading order analysis given in \S\ref{sec:EquivBCs}. The higher-order expansion is required for our analysis of near-resonance effects in the Dirichlet case, and the `small gap' regime in the Neumann case. 
We begin by noting that the two-term boundary layer equation generalising \rf{eqn:innerOneTerm} is
\begin{align}
\label{eqn:BLPDE}
\pdtwo{\Phi}{N} + \pdtwo{\Phi}{S} + \eps \kappa \left(\pdone{\Phi}{N} - 2 N\pdtwo{\Phi}{S}\right) + 2\eps \frac{\partial^2 \Phi}{\partial S\partial s} + \ord{\eps^2} & = 0.
\end{align} 
and the periodicity condition remains (\ref{eqn:periodicity}).
The cell domain on which \rf{eqn:BLPDE} is to hold is different for the two wire models.  For Model 2, when the wire shape is defined in the curvilinear coordinates, it is simply $\mathcal{B}=\{(N,S):|S|<1/2\}\setminus \mathcal{K}$, where $\mathcal{K} = \delta K$, and homogenous Dirichlet or Neumann boundary conditions (as appropriate) are imposed on the scaled wire boundary $\partial \mathcal{K}$.  

For Model 1, the curvature of $\Gamma$ complicates matters somewhat.  A priori the domain is $\mathcal{B}_\eps=\{(N,S):|S|<1/2\}\setminus \mathcal{K}_\eps$, where $\mathcal{K}_\eps$ is the scaled wire described in $(N,S)$ coordinates, and the boundary conditions  are to be imposed on $\mathcal{K}_\eps$.  As illustrated in Figure \ref{fig:geometryMOD}, $\mathcal{K}_\eps$ is in general perturbed from $\mathcal{K}$, depending on $\eps$ and the local curvature of $\Gamma$. This is undesirable, and it is preferable to solve cell problems on the fixed cell domain $\mathcal{B}=\{(N,S):|S|<1/2\}\setminus \mathcal{K}$, as for Model 2.  If we only desire the leading order approximation of $\Phi(N,S;s)$, as in the main text, the change from $\mathcal{B}_\eps$ to $\mathcal{B}$ incurs no loss of asymptotic accuracy.  But when higher order terms are required, one needs to expand the boundary conditions carefully so as to compensate for the geometric deformation. 

To do this, note that the relationship $(\tilde{x},\tilde{y}) = F_j(\tilde{n},\tilde{s})$ between the local curvilinear and Cartesian coordinates, introduced in \S\ref{sec:ProblemFormulation}, can be written in terms of the boundary-layer coordinates $(N,S)$ as
\begin{align}
\label{eq:coordinatechange}
\tilde{x}/\eps = N - \half \eps \kappa S^2 +  \ord{\eps^2}, \qquad
\tilde{y}/\eps = S + \eps \kappa N S + \ord{\eps^2},
\end{align}
as $\eps \to 0$, where $\kappa = \kappa(s)$ is the local curvature of $\Gamma$.  We suppose for definiteness that the boundary of the reference wire shape, $\partial K$, is smooth and is given by $W(\xi,\eta) = 0$ in Cartesian coordinates $(\xi,\eta)$ (cf.\ Figure \ref{fig:geometryMOD}\subref{subfig:K}).  The boundary $\partial \mathcal{K}$ is then given by $W(N/\delta,S/\delta)=0$ (this is the actual wire boundary under Model 2), while $\partial \mathcal{K}_{\eps}$ is given by $W(\tilde{x}/\delta\eps,\tilde{y}/\delta\eps) = 0$. 
Expanding this expression using (\ref{eq:coordinatechange}) shows that the Dirichlet condition $\Phi(N,S) = 0$ on $\partial {\cal K}_{\eps}$ can be replaced by
\begin{align}
\label{eqn:BLDirichlet}
\Phi(N,S) + \eps \kappa\,d\, \pdone{\Phi}{\nu}(N,S)  + \ord{\eps^2} = 0 \qquad  \textrm{on } \partial {\cal K},
\end{align}
where $\kappa(s)\, d(N,S)$ is the normal perturbation of $\partial {\cal K}_{\eps}$ from $\partial {\cal K}$, with
\begin{align}
d(N,S) = (\half S^2, -NS) \cdot \nu,
\end{align}
where $\nu=(\nu_N,\nu_S) = \nabla W / |\nabla W|$ is the outward unit normal to $\partial {\cal K}$.  

A more involved calculation shows that the Neumann condition $\pdonetext{\Phi}{\nu}(N,S)=0$ on $\partial\mathcal{K}_\eps$ can be replaced by
\begin{align}
\label{eqn:BLNeumann}
\pdone{\Phi}{\nu}(N,S) + \eps\kappa \Big(d\,\pdtwo{\Phi}{\nu}(N,S)
+ \tilde{d}\,\pdone{\Phi}{\nu^\perp}(N,S)\Big) +\ord{\eps^2} = 0 \qquad \textrm{on } \partial\mathcal{K},
\end{align}
with 
\begin{align}
\tilde{d}(N,S) = -S + N \nu_N \nu_S - \kappa_{\partial\mathcal{K}}(N,S)\, (\half S^2, -NS) \cdot \nu^\perp,
\end{align}
where $\nu^\perp=(-\nu_S,\nu_N)$ is the (counterclockwise) unit tangent vector on $\partial\mathcal{K}$, and 
$\kappa_{\partial\mathcal{K}}(N,S)$ is the curvature of $\partial\mathcal{K}$ at the point $(N,S)$. (The normal to $\partial\mathcal{K}_{\eps}$ is given by $\nu + \eps \kappa\, \tilde{d}(N,S) \nu^\perp + \ord{\eps^2}$.)

As a concrete example, consider the circular disc, when $W(\xi,\eta) = \xi^2 + \eta^2 - 1$. Parameterising $\partial\mathcal{K}$ by $(N,S) = \delta (\cos\vartheta,\sin\vartheta)$ for $\vartheta\in[0,2\pi)$ gives $\nu = (\cos\vartheta,\sin\vartheta)$, $\kappa_{\partial\mathcal{K}} = 1/\delta$, $d  = - \half \delta^2 \cos\vartheta \sin^2\vartheta $, and $\tilde{d} = \delta \sin\vartheta (1-\mbox{$\frac{3}{2}$} \sin^2\vartheta)$.

To summarise, the boundary-layer problems for Model 1 are given by (\ref{eqn:BLPDE}) with periodic boundary conditions (\ref{eqn:periodicity}), one of (\ref{eqn:BLDirichlet}) or (\ref{eqn:BLNeumann}), and matching conditions as $N \to \pm \infty$.  The problems for Model 2 are the same except that the geometric correction terms involving $d$ and $\tilde{d}$ in (\ref{eqn:BLDirichlet}) and (\ref{eqn:BLNeumann}) are ignored.  We proceed with the analysis for Model 1, but note that the corresponding solutions for Model 2 can be obtained simply by setting $d = \tilde{d} = 0$ in the following.

\subsection{Dirichlet problem}

For the Dirichlet problem the leading order solution has the general form
\begin{align}
\Phi_0(N,S;s) = A_0^+(s) \Phi^+(N,S) + A_0^-(s) \Phi^-(N,S), 
\end{align}
where $\Phi^+$ and $\Phi^-$ are the canonical solutions defined earlier in (\ref{eqn:cell1})-(\ref{eqn:cell2}).

The $\ord{\eps}$ solution can be written as
\begin{align}
\Phi_1(N,S;s) = A_1^+(s) \Phi^+ + A_1^-(s) \Phi^- + \kappa(s) A_0^+  \tilde{\Phi}^+ + \kappa(s) A_0^- \tilde{\Phi}^- + \pdone{A_0^+}{s} \hat{\Phi}^+ + \pdone{A_0^-}{s} \hat{\Phi}^-, 
\end{align}
where $\tilde{\Phi}^{\pm}$ and $\hat{\Phi}^{\pm}$ satisfy canonical problems defined in terms of $\Phi^\pm$.  The problem for $\tilde{\Phi}^{\pm}$ is
\begin{align}
\pdtwo{\tilde{\Phi}^{\pm}}{N} + \pdtwo{\tilde{\Phi}^{\pm}}{S}  = - 2N\pdtwo{\Phi^{\pm}}{N} - \pdone{\Phi^{\pm}}{N} &\qquad \textrm{in }\mathcal{B} \\
\tilde{\Phi}^{\pm}(N,-1/2)=\tilde{\Phi}^{\pm}(N,1/2),  &\\
 \tilde{\Phi}^{\pm} = - d(N,S) \pdone{\Phi^{\pm}}{\nu} &\qquad  \textrm{on } \partial\mathcal{K},\\
\tilde{\Phi}^+(N,S)\sim
\begin{cases}
- \half N^2 + \tilde{\sigma}_+, & N\to\infty,\\
\tilde{\tau}_+, & N\to-\infty,\\
\end{cases}
& \qquad
\tilde{\Phi}^-(N,S)\sim
\begin{cases}
-\tilde{\tau}_-, & N\to\infty,\\
\half N^2 -\tilde{\sigma}_-, & N\to-\infty, \label{eqn:sigmatilde}\\
\end{cases}
\end{align} 
where the constants $\tilde{\sigma}_{\pm}$ and $\tilde{\tau}_{\pm}$ are determined as part of the solution.  If the wire shape $K$ is symmetric in $\xi$, then from \rf{eqn:PhiPMSymmetry} it follows that $\tilde{\Phi}^+(N,S) = - \tilde{\Phi}^-(-N,S)$, so $\tilde{\tau}_- =  \tilde{\tau}_+$, $\tilde{\sigma}_- =  \tilde{\sigma}_+$.

The problem for $\hat{\Phi}^{\pm}$ is
\begin{align}
\pdtwo{\hat{\Phi}^{\pm}}{N} + \pdtwo{\hat{\Phi}^{\pm}}{S}  = - 2\pdone{\Phi^{\pm}}{S}  
&\qquad \textrm{in }\mathcal{B}, \\
\hat{\Phi}^{\pm}(N,-1/2)=\hat{\Phi}^{\pm}(N,1/2),  &\\
 \hat{\Phi}^{\pm} = 0 &\qquad  \textrm{on } \partial\mathcal{K},\\
\hat{\Phi}^+(N,S)\sim
\begin{cases}
\hat{\sigma}_+, & N\to\infty,\\
\hat{\tau}_+, & N\to-\infty,\\
\end{cases}
& \qquad
\hat{\Phi}^-(N,S)\sim
\begin{cases}
\hat{\tau}_-, & N\to\infty,\\
\hat{\sigma}_-, & N\to-\infty,\\
\end{cases}
\end{align}
where the constants $\hat{\sigma}_{\pm}$ and $\hat{\tau}_{\pm}$ are again determined as part of the solution. If $K$ is symmetric in $\xi$ then $\hat{\Phi}^+(N,S) = \hat{\Phi}^-(-N,S)$, so $\hat{\tau}_- = \hat{\tau}_+$ and $\hat{\sigma}_- = \hat{\sigma}_+$.

The far-field behaviour as $N \to \pm \infty$ of the two-term solution \rf{eqn:BLTwoTerm}, required for matching, is then
\begin{align}
\label{eqn:FFInner}
\Phi \sim &   \mp \half \eps \kappa A_0^{\pm} N^2 \pm (A_0^\pm + \eps A_1^\pm ) N  + A_0^\pm \sigma_\pm + A_0^\mp \tau_\mp \nonumber\\ &+  \eps A_1^\pm \sigma_\pm + \eps A_1^\mp \tau_\mp \pm \eps \kappa A_0^\pm  \tilde{\sigma}_\pm \mp \eps \kappa A_0^\mp \tilde{\tau}_\mp + \eps \pdone{A_0^\pm}{s} \hat{\sigma}_\pm + \eps \pdone{A_0^\mp}{s} \hat{\tau}_\mp + \ord{\eps^2}. %
\end{align}

\subsection{Neumann problem}

For the Neumann problem the $\ord{\eps}$ solution can be written as
\begin{align}
\Phi_0(N,S;s) + \eps \Phi_1(N,S;s) = A_0(s) + \eps \Big( A_1(s) + B_1(s) \Psi (N,S)\Big), 
\end{align}
where $\Psi$ is the solution of the canonical problem defined in (\ref{eqn:Ncell1})-(\ref{eqn:Ncell2}).

The $\ord{\eps^2}$ solution can be written as
\begin{align}
\Phi_2(N,S;s) = A_2(s) + B_2(s) \Psi  + \kappa(s) B_1 \tilde{\Psi}  + \pdone{B_1}{s} \hat{\Psi} +  \left( \pdtwo{A_0}{s} + k^2 A_0 \right) \check{\Psi},
\end{align}
where $\tilde{\Psi}(N,S)$ and $\hat{\Psi}(N,S)$ satisfy canonical problems involving $\Psi(N,S)$, and $\check{\Psi}(N,S)$ satisfies a canonical correction problem arising from the constant solution.  The problem for $\tilde{\Psi}(N,S)$ is
\begin{align}
\label{eqn:tildePsi1}
\pdtwo{\tilde{\Psi}}{N} + \pdtwo{\tilde{\Psi}}{S}  = - 2N\pdtwo{\Psi}{N} - \pdone{\Psi}{N} &\qquad \textrm{in }\mathcal{B} \\
\tilde{\Psi}(N,-1/2)=\tilde{\Psi}(N,1/2),  &\\
\pdone{\tilde{\Psi}}{\nu} = - d(N,S) \pdtwo{\Psi}{\nu}- \tilde{d}(N,S) \pdone{\Psi}{\nu^\perp} &\qquad \textrm{on } \partial\mathcal{K},\\
\label{eqn:tildePsi2}
\tilde{\Psi}(N,S)\sim
- \half N^2 \pm \tilde{\mu}N \pm \tilde{\lambda}, &\qquad N\to \pm\infty,
\end{align}
where the constants $\tilde{\mu}$ and $\tilde{\lambda}$ are determined as part of the solution.  If $K$ is symmetric in $\xi$ then $\Psi(N,S)=-\Psi(-N,S)$, so that $\tilde{\Psi}(N,S)=\tilde{\Psi}(-N,S)$ and $\tilde{\lambda} = 0$. 

The constant $\tilde{\mu}$ arising here can be evaluated directly from $\Psi$, by integrating (\ref{eqn:tildePsi1}) over $\mathcal{B}$ and using the divergence theorem,
which yields, after application of the boundary and matching conditions, the formula
\begin{align}
\label{eqn:tildemuformula}
\tilde{\mu} = \lambda - \frac{1}{2}\int_{\partial\mathcal{K}} \left[ \left(\Psi - 2N\pdone{\Psi}{N} \right) \nu_N  + d(N,S)\, \pdtwo{\Psi}{\nu} + \tilde{d}(N,S)\, \pdone{\Psi}{\nu^{\perp}}\right] {\,\rm d}\nu^{\perp}.
\end{align}

The problem for $\hat{\Psi}(N,S)$ is
\begin{align}
\pdtwo{\hat{\Psi}}{N} + \pdtwo{\hat{\Psi}}{S}  = - 2\pdone{\Psi}{S}  
&\qquad \textrm{in }\mathcal{B}, \\
\hat{\Psi}(N,-1/2)=\hat{\Psi}(N,1/2),  &\\
\pdone{\hat{\Psi}}{\nu} = 0 &\qquad \textrm{on } \partial\mathcal{K},\\
\hat{\Psi}(N,S) \sim
\pm \hat{\mu} N \pm\hat{\lambda}, &\qquad  N\to\pm\infty,
\end{align}
where, again, the constants $\hat{\mu}$ and $\hat{\lambda}$ are determined as part of the solution. If $K$ is symmetric in $\xi$ then $\hat{\Psi}(N,S)=-\hat{\Psi}(-N,S)$ and $\hat{\mu} = 0$.  

Finally, the problem for $\check{\Psi}(N,S)$ is
\begin{align}
\label{eqn:checkPsi1}
\pdtwo{\check{\Psi}}{N} + \pdtwo{\check{\Psi}}{S}  = - 1 
&\qquad \textrm{in }\mathcal{B}, \\
\check{\Psi}(N,-1/2)=\check{\Psi}(N,1/2), &\\
\pdone{\check{\Psi}}{\nu} = 0 &\qquad \textrm{on } \partial\mathcal{K},\\
\check{\Psi}(N,S) \sim
-\half N^2 \pm \check{\mu} N \pm\check{\lambda}, &\qquad  N\to\pm\infty,
\end{align}
If $K$ is symmetric in $\xi$ then $\check{\Psi}(N,S)=\check{\Psi}(-N,S)$ and $\check{\lambda} = 0$.   Integrating (\ref{eqn:checkPsi1}) over $\mathcal{B}$, as before, yields the formula $\check{\mu} = \half {\rm Area}(\mathcal{K})$ \cite{MaDa:88}.

The far-field behaviour as $N \to \pm \infty$ of \rf{eqn:BLTwoTerm} is then
\begin{align}
\label{eqn:FFInnerNeumann}
\Phi \sim &   -\half \eps^2 \kappa B_1 N^2 -\half \eps^2 \left( \pdtwo{A_0}{s} + k^2 A_0 \right)  N^2  
\nonumber \\ &
+ \left(\eps B_1  + \eps^2 B_2 \pm \eps^2 \kappa \tilde{\mu} B_1  \pm \eps^2 \pdone{B_1}{s} \hat{\mu} \pm \eps^2 \left( \pdtwo{A_0}{s} + k^2 A_0 \right) \check{\mu} \right)N 
\nonumber \\  &
+ A_0 + \eps A_1 \pm \eps B_1 \lambda  + \eps^2 A_2 \pm  \eps^2 B_2 \lambda \pm \eps^2 \kappa \tilde{\lambda} B_1 \pm \eps^2  \pdone{B_1}{s} \hat{\lambda} \pm \eps^2  \left( \pdtwo{A_0}{s} + k^2 A_0 \right) \check{\lambda} + \ord{\eps^3} . %
\end{align}
\section{Cell problem solutions\label{sec:cells}}
In this section we present numerical and analytical solutions to the leading order boundary layer cell problems for disk-shaped, perpendicular/tangential line segments, and square wires.

\subsection{Dirichlet problems}

\begin{figure}[t]
\centering
\includegraphics[width=\textwidth]{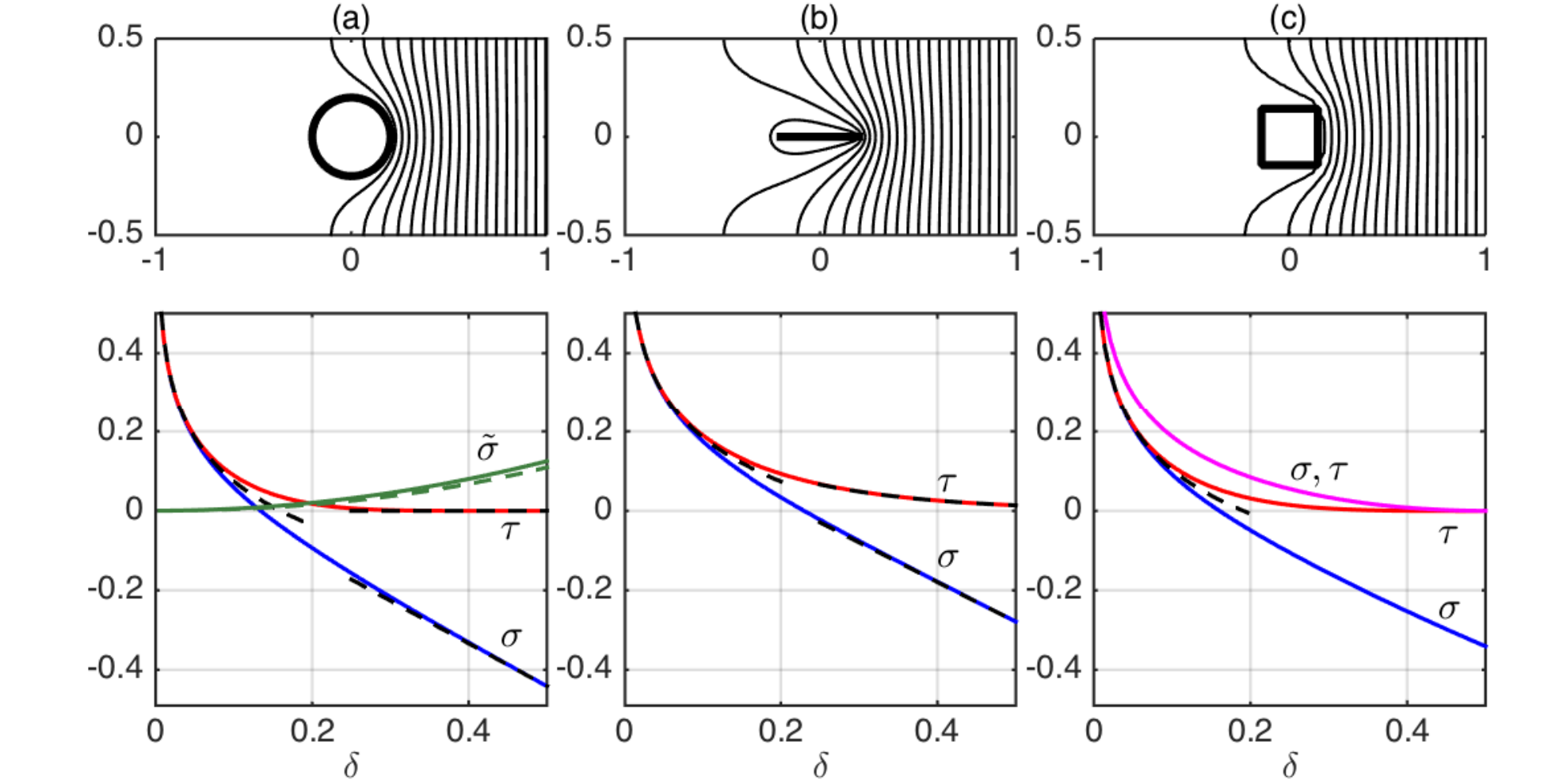}
\caption{\label{fig:inner1}Solutions to the cell problem (\ref{eqn:cell1})-(\ref{eqn:cell2}) for (a) disk-shaped wires with radius $\delta$, (b) infinitely thin perpendicular line segments with length $2\delta$, and (c) square wires with side length $\sqrt{2}\delta$.   Upper panels show contours of $\Phi^+(N,S)$ for $\delta = 0.2$. Lower panels show the constants $\sigma$ and $\tau$ in the far field expansion, as a function of $\delta$.  
Dashed lines show asymptotic behaviour of $\sigma$ and $\tau$ as $\delta$ approaches $0$ and $\half$ (or $\delta \to \infty$ in (b)).  In (a), the higher-order correction $\tilde{\sigma}$ from (\ref{eqn:sigmatilde}) is also shown, for Model 1 (solid) and Model 2 (dashed), and in (c) the magenta curve shows $\sigma = \tau$ for a tangential line segment of length $2\delta$, from (\ref{eqn:tangentialsigmatau}).}
\end{figure}

Example solutions for the Dirichlet cell problem \rf{eqn:cell1}-\rf{eqn:cell2} are shown in Figure \ref{fig:inner1}, along with plots of the corresponding far-field constants $\sigma=\sigma^\pm$ and $\tau=\tau^\pm$. 

The solutions for circular wires in Figure \ref{fig:inner1}(a) are calculated numerically using linear finite elements, with the constants $\sigma$ and $\tau$ found from a linear fit of the far-field behaviour.  For small wires, $\delta \to 0$, we recall the asymptotic behaviour \rf{eqn:BLSolnDeltaSmall}-\rf{eqn:MuNuDeltaSmall}. 
For $\delta = \deltamax=\half$, the circle takes up the whole width of the cell domain and $\Phi^+ = 0$ for $X < 0$ (so $\tau = 0$), while a numerical solution gives $\sigma \approx -0.44 $ for the constant as $X \to \infty$. 
As $\delta \to \half$, the asymptotic behaviour of $\sigma$ can be found by solving a perturbation problem numerically, from which we obtain $\sigma \sim -0.44 + 1.07 (\half-\delta)$, while $\tau$ is exponentially small. The solutions for the square wire in Figure \ref{fig:inner1}(c) were also computed numerically; we note that in this case $\delta_{\rm max}=1/\sqrt{2}$.

Solutions for the two arrangements of line segments can be found analytically by conformal mapping.  For the wires arranged perpendicular to $\Gamma$ we obtain
\begin{align}
\label{}
\Phi^+(N,S) =\Re \left\{ \frac{1}{2\pi} \log \left[\frac{e^{-\pi \delta} + \zeta}{e^{-\pi \delta} -\zeta} \right] \right\}, \quad \zeta = \left[ \frac{\sinh \pi (Z-\delta) }{\sinh \pi (Z+\delta)} \right]^{1/2},
\quad Z=N+\ri S,
\end{align}
from which we find
\begin{align}
\label{}
\sigma = -\frac{1}{2\pi} \log\left( \frac{\sinh 2\pi \delta}{2} \right), \quad
\tau = - \frac{1}{2\pi} \log\left( \tanh\pi \delta \right).
\end{align}
These have limiting behaviour $\sigma \sim \tau \sim -(1/2\pi) \log(\pi \delta)$ as $\delta \to 0$ (so that in particular $a_0=\log{2}$), and $\sigma \sim -\delta + (1/\pi) \log 2 $, $\tau \sim (1/\pi)e^{-2\pi \delta} $ as $\delta \to \infty$, as shown in Figure \ref{fig:inner1}(b).

For the wires arranged tangentially along $\Gamma$ we obtain
\begin{align}
\label{}
\Phi^+(N,S) = \Re \left\{ \frac{1}{2\pi} \log \left[\frac{e^{i \pi \delta} + \zeta}{e^{- i \pi \delta} - \zeta} \right] \right\} , \quad \zeta = \left[ \frac{\sin \pi (i Z + \delta) }{\sin \pi (i Z- \delta)} \right]^{1/2},
\quad Z=N+\ri S,
\end{align}
from which we find
\begin{align}
\label{eqn:tangentialsigmatau}
\sigma = \tau = -\frac{1}{2\pi} \log\left( \sin\pi \delta \right). \qquad
\end{align}
This again has $\sigma \sim \tau -(1/2\pi) \log(\pi \delta)$ as $\delta \to 0$ (so $a_0=\log2$), 
while $\sigma \sim \tau \sim \mbox{$\frac{1}{4}$} \pi (\half-\delta)^2$ as $\delta \to \half$.

\subsection{Neumann problems}

\begin{figure}[t]
\centering
\includegraphics[width=\textwidth]{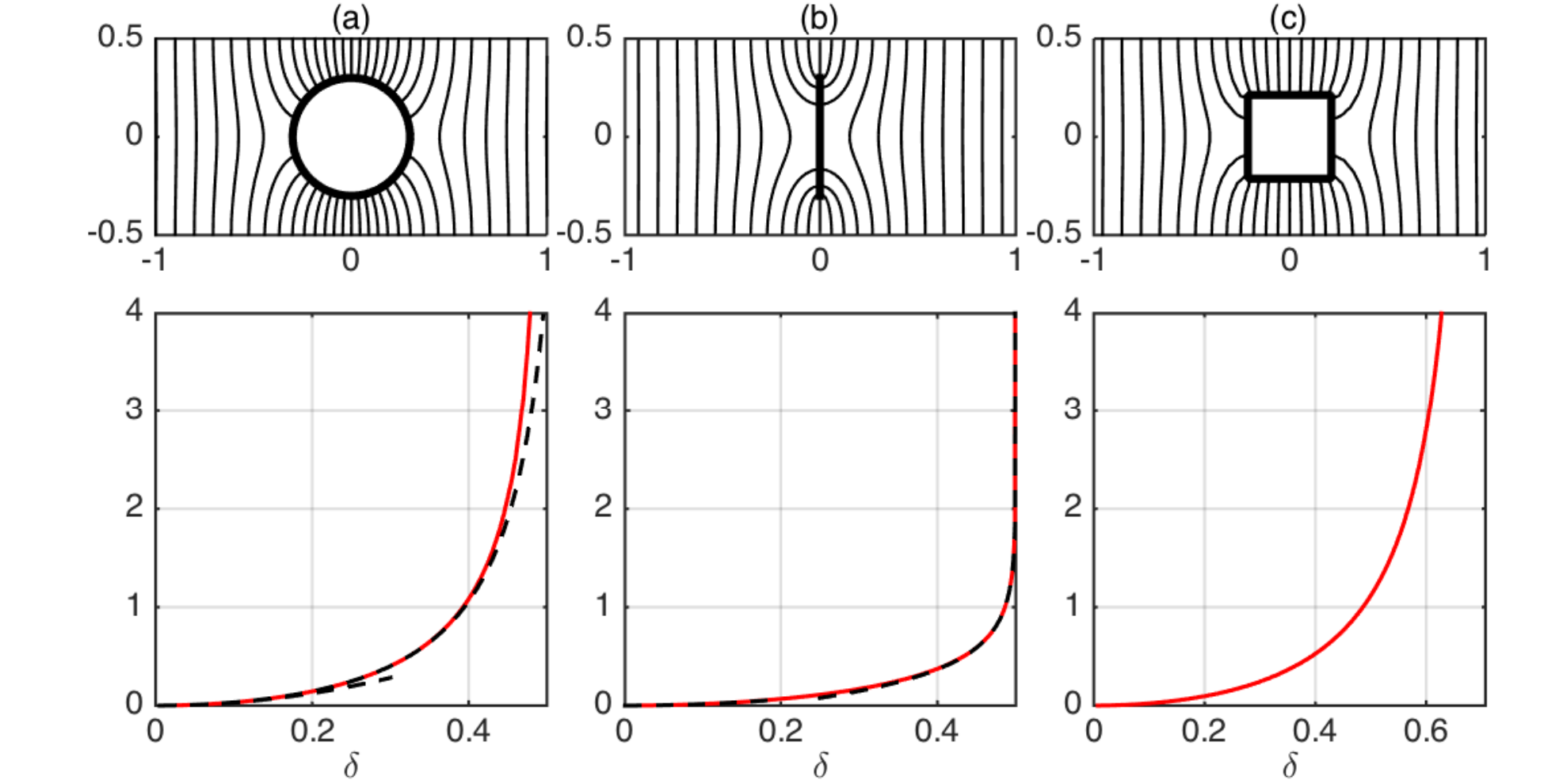}
\caption{\label{fig:inner2}Solutions to the Neumann cell problem (\ref{eqn:Ncell1})-(\ref{eqn:Ncell2}) for (a) disk-shaped wires with radius $\delta$, (b) infinitely thin line segments with length $2\delta$, and (c) square wires with side length $\sqrt{2}\delta$. Upper panels show contours of $\Psi(N,S)$ for $\delta = 0.3$.  Lower panels show the constant $\lambda$ in the far field expansion, as a function of $\delta$.  The dashed lines in (a) show the approximations $\lambda \sim \pi \delta^2$ and $\lambda \sim \pi\delta^2/(1-(\pi\delta)^2/3)$ \cite{Cr:16}, and in (b) show the behaviour as $\delta$ approaches $0$ and $\half$.
}
\end{figure}

Example solutions for the Neumann cell problem (\ref{eqn:Ncell1})-(\ref{eqn:Ncell2}) are shown in Figure \ref{fig:inner2}.  

The circular wire case is again calculated numerically, although the asymptotic behaviour for small and large circles provides a good fit over the whole range of $\delta$.  For $\delta \to 0$, the solution away from the wire can be written approximately as
\begin{align}
\label{}
\Psi(N,S) \sim \Re \left\{ Z + \frac{\delta^2 \pi}{\tanh \pi Z} \right\},\qquad Z=N+\ri S,
\end{align}
which gives $\lambda \sim \pi \delta^2$ as $\delta \to 0$. (The strength of the singularity here is again determined by matching to an inner region close to the wire, as in \cite[\S B]{ChHeTr:15}, where $\Psi \sim \Re \left\{ Z + \delta^2 /Z \right\}$). 
We remark that the analysis in \cite{Cr:16} provides a more refined approximation $\lambda\sim (\pi\delta^2)/(1-(\pi\delta)^2/3)$, which is also plotted in Figure \ref{fig:inner2}.
For $\delta \to \half \pi$, one can show that $\lambda \sim \mbox{$\frac{1}{4}$} \pi (\half - \delta)^{-1/2}$.

For line segments arranged perpendicular to $\Gamma$, the wire has no impact on the solution, which is simply $\Psi(N,S) = N$, so $\lambda = 0$.   For line segments arranged tangentially along $\Gamma$, conformal mapping yields
\begin{align}
\label{eqn:neumannpsisolution}
\Psi(N,S) = \Re \left\{ \frac{1}{2\pi} \log \left[\frac{(e^{i \pi \delta} - \zeta)(e^{-i \pi \delta} + \zeta)}{(e^{-i \pi \delta} - \zeta)(e^{i \pi \delta} + \zeta)} \right] \right\} , \quad \zeta = \left[ \frac{\sin \pi (i Z + \delta) }{\sin \pi (i Z- \delta)} \right]^{1/2}, 
\quad Z=N+\ri S,
\end{align}
from which we find
\begin{align}
\label{}
\lambda = -\frac{1}{\pi} \log\left( \cos \pi \delta \right). \qquad
\end{align}
This has limiting behaviour $\lambda \sim \half \pi \delta^2$ as $\delta \to 0$, and $\lambda \sim -(1/\pi) \log \pi (\half -\delta)$ as $\delta \to \half$.
Using (\ref{eqn:neumannpsisolution}) in (\ref{eqn:tildemuformula}), for Model 2, we find that $\tilde{\mu} = 0$, along with $\hat{\mu} = 0$ (by symmetry) and $\check{\mu} = 0$.

\end{document}